\documentclass[reprint,amstrange metalath,aps,notitlepage,superscriptaddress,floatfix,pra]{revtex4-2}

\usepackage{placeins}
\usepackage{graphicx,epstopdf,dcolumn,bm,mathtools,siunitx,multirow,xcolor,amsfonts,physics,bibunits}

\usepackage{graphicx}
\usepackage{dcolumn}
\usepackage{bm}

\begin{document}                                                                                                                                                                                                                                                                                                                                                                             \title{Quantum Acoustics Demystifies the  Strange Metals}
                                                                                                                                                                                                                                                                                                                                                                                                                                                        
\author{Eric J.~Heller} 
\affiliation{Department of Physics, Harvard University, Harvard University, Cambridge, MA 02138, USA}
\affiliation{Department of Chemistry and Chemical Biology, Harvard University, Cambridge,
MA 02138, USA}
\author{Alhun~Aydin}
\affiliation{Department of Physics, Harvard University, Harvard University, Cambridge, MA 02138, USA}
\affiliation{Faculty of Engineering and Natural Sciences, Sabanci University, 34956 Tuzla, Istanbul, Turkey}

\author{Anton M.~Graf}
\affiliation{Harvard John A. Paulson School of Engineering and Applied Sciences,
Harvard, Cambridge, Massachusetts 02138, USA}
\affiliation{Department of Chemistry and Chemical Biology, Harvard University, Cambridge,
MA 02138, USA}
\author{Yubo Zhang}\affiliation{School of Physics, Peking University, No.5 Yiheyuan Rd, Beijing
100871, China}

\author{Joost de Nijs}

\affiliation{Faculty of Applied Sciences, Delft University of Technology, 2628 CD Delft, Netherlands}
\affiliation{Institute of Applied Mathematics, Delft University of Technology, 2628 CD Delft, Netherlands}                                                                                                                                                                                                              
\author{Yoel Zimmermann} \affiliation{ Department of Chemistry and Applied Biosciences, ETH Zurich, 8093 Zurich, Switzerland}

\author{Xiaoyu~Ouyang}
\affiliation{Yuanpei College, Peking University, No.5 Yiheyuan Rd, Beijing
100871, China}
\affiliation{School of Physics, Peking University, No.5 Yiheyuan Rd, Beijing
100871, China}
\author{Shaobing~Yuan}
\affiliation{School of Physics, Peking University, No.5 Yiheyuan Rd, Beijing
100871, China}
\author{Alvar Daza}
\affiliation{Nonlinear Dynamics, Chaos and Complex Systems Group, Departamento de F\'isica, Universidad Rey Juan Carlos,
Tulip\'an s/n, 28933 M\'ostoles, Madrid, Spain}
\affiliation{Department of Physics, Harvard University, Harvard University, Cambridge, MA 02138, USA}

\author{Zixuan Chai}
\email{zc362@cam.ac.uk}
\affiliation{%
 Department of Physics, University of Cambridge, 
}%
\affiliation{%
 Department of Physics, Harvard University, Cambridge, Massachusetts 02138, USA
}%
\author{Siyuan Chen}
\email{alaincsy@stu.pku.edu.cn  }
\affiliation{%
 Department of Physics, Harvard University, Cambridge, Massachusetts 02138, USA
}%
\author{Jasper Jain }
\email{jasperjain@college.harvard.edu}
\affiliation{
 Department of Physics, Harvard University, Cambridge, Massachusetts 02138, USA
}%
\author{Mingxuan Xiao }
\email{xiaomx@stu.pku.edu.cn}
\affiliation{%
 Department of Physics, Harvard University, Cambridge, Massachusetts 02138, USA
}%

\author{Chenzheng Yu}\affiliation{School of Physics, Peking University, No.5 Yiheyuan Rd, Beijing
100871, China}
\author{Zhongling Lu}
\affiliation{Department of Chemistry and Chemical Biology, Harvard University, Cambridge,
MA 02138, USA}

 \author{Joonas~Keski-Rahkonen}
\affiliation{Department of Physics, Harvard University, Harvard University, Cambridge, MA 02138, USA}

\date{\today} ~
\begin{abstract}
Phonons have long been thought to be incapable of explaining key phenomena in strange metals, including linear-in-\textit{T} Planckian resistivity from high to very low temperatures.  We argue that these conclusions were based on static,  perturbative approaches that overlooked essential time-dependent and nonperturbative electron-lattice physics. In fact ``phonons'' are not the best target for discussion, just like ``photons'' are not the best way to think about Maxwell's equations. Quantum optics connects photons and electromagnetism, as developed 60 years ago by Glauber and others. We have developed a parallel world of quantum acoustics. Far from being only of academic interest, the new tools are rapidly exposing the secrets of the strange metals, revealing strong vibronic (vibration-electronic) interactions playing a crucial role forming polarons and charge density waves, linear-in-$T$  resistivity at the Planckian rate over thousands of degrees, resolution of the Drude peak infrared anomaly, and the absence of a  $T^4$  low-temperature resistivity rise in 2D systems, and of a Mott-Ioffe-Regel resistivity saturation. We derive Planckian transport, polarons, CDWs, and pseudogaps from the Fr\"ohlich
model. The ``new physics'' has been hiding in this model all along, in the right parameter regime, if it is treated  nonperturbatively. In the course of this work we have uncovered the generalization of Anderson localization to dynamic media: a universal Planckian diffusion emerges, a ``ghost'' of Anderson localization. Planckian diffusion is clearly defined and is more fundamental than the popular but elusive,  model dependent concept of  ``Planckian speed limit''.  
   
\end{abstract}
\maketitle
\tableofcontents


\section{Introduction}\label{s:intro}

In the 1950s and 1960s, the Hanbury-Brown Twiss experiment made it imperative to unify the particle-like photon picture of light with the wave-like electromagnetic field obeying Maxwell's equations. Thus was born the new paradigm of quantum optics, where Glauber coherent states became the vehicle to travel from photons to electromagnetic waves.    

The  opportunity to develop solid state theory of lattice vibration along parallel lines has been largely neglected. The phonon plays the role of the photon, and the acoustic waves play the role of the electromagnetic waves.

The  connection goes back to Schr\"odinger, who worked out the number state-coherent state connection in 1926, Fig. \ref{Schroedinger1926.jpg} \cite{schr}.  The former is natural to the time-independent Schr\"odinger equation, and the latter needs the time-dependent Schr\"odinger equation.  

We have recently been developing this paradigm for electrons in a lattice and  have found it to be remarkably powerful for both physical insight and computations. We call it quantum acoustics. We do not need to introduce a new physical model, but rather we develop the Fr\"ohlich model in a perfectly valid but neglected coherent state representation.  New physics emerges, as well-grounded as it is in quantum optics.

The traditional solid state paradigm is one of particles, i.e. phonons.  In special contexts, involving transducers, for example,  acoustic waves are familiar, but the unification of phonons and acoustic waves has been  lacking. It is natural that the coherent state is again the right tool to unify the  wave and particle pictures, this time in solid state physics.  Far from being of only academic interest, quantum acoustics has made it possible to understand several of the high-temperature superconductor, above-$T_c$ ``strange metal'' mysteries, within a single theoretical context and within perhaps the most well established model of electron-vibration interactions (Fr\"ohlich). This paper reviews that progress and  points the way to  likely future developments.


Earlier, we put the time-dependent real space ``wave-on-wave,'' or WoW code described below  through its paces regarding ordinary metals~\cite{heller22}. It gave essentially exact agreement with the Bloch-Gr\"uneisen, perturbative + Boltzmann transport theory, including a $T^4$ ($T^5$ in 3D) low temperature rise of resistivity at low $T$~\cite{heller22}. Neither perturbation theory nor Boltzmann transport was used. Now we are applying quantum acoustics to the high $T_c$ materials in their strange ``normal'' (non-superconducting) state, and it is yielding a cornucopia of  results. The WoW approach, including the lattice interacting with the electrons and {\it vise-versa} is nonperturbative and coherence preserving; both traits are needed to understand the strange metals. 

 We have already explained or believe  we will be able to explain most of the above experimental results above T$_c$, but we cannot yet claim to have pinned down the very low temperature linearity of resistivity. We   have plausible lattice-electron scenarios, but this leads us into speculation, which we mostly avoid here. 


 The veil is lifting on   many of the strange metal mysteries and controversies, after nearly 40 years. It has become crystal clear that electron-lattice interactions are  essential to understand  the strange metal physics.  

``Strongly correlated'' behavior is often cited as a hallmark of strange metals. In our perspective, there is a strong coupling that emerges within the venerable Fr\"ohlich Hamiltonian, and  is  {\it vibronic} in origin: 
Electrons become strongly coupled indirectly with each other because they are each strongly coupled to the same lattice and its deformations due to thermal vibrations and the presence of the electron back-action. Even near 0 Kelvin, given a deformation potential constant of 10 or 20 eV,   strong electron-lattice coupling emerges in strange metals (but not for ordinary metals).     

The  basis for this approach can be traced back to the semiclassical (with respect to the lattice)  polaron theory of Landau and Pekar~\cite{landau1948effective}, as amplified by Leopold {\it et al.} ~\cite{leopold2022landau}.  In analogy with an electron in a classical electromagnetic field, Landau-Pekar takes a classical lattice approach as opposed to a quantized, number-state, phonon-based approach.   

Another important antecedent is Peierls distortion of a lattice.  This very idea shifts us quietly into a coherent state representation of the lattice because it is extremely inconvenient to think of the distortions in terms of phonon occupations. Sound waves in a crystal coming from a transducer also necessarily slip us into the coherent state representation, or at least a classical lattice picture. 


The Fr\"ohlich Hamiltonian, with its centerpiece electron-phonon interaction term, generates strongly coupled emergent behavior, if it is allowed to remain always ``on,'' acting coherently to unlimited order. The quantum acoustics paradigm can implement the Fr\"ohlich Hamiltonian nicely, but it should be noted that the quantum acoustics framework is much more general than the Fr\"ohlich model.  That is, much higher order models can also be accommodated with quantum acoustics. 

We make the case below that quantum acoustics, wherein ``coherent and always on'' electron lattice interactions happen naturally,  holds the key to much of the strange metal phenomenology. Several papers have laid the groundwork~\cite{heller22,PhysRevLett.132,Aydin24,aydin_polaron_2025} and have offered plausible explanations to important strange metal mysteries. 


Using quantum acoustics framework, we have so far provided explanations for (a) bypass of the Mott-Ioffe-Regel limit~\cite{heller22}, (b) low-frequency Drude peak suppression and displacement~\cite{PhysRevLett.132}, (c) linear-in-\textit{T} resistivity with Planckian slope from hundreds of degrees down to at least 50K, and lower with strong magnetic fields~\cite{Aydin24}, and (d) definitive and spontaneous ``real time'' polaron and CDW formation in dynamical WoW simulations~\cite{aydin_polaron_2025}, superseding linear response and perturbative approaches. 

The term ``quantum acoustics'' is not new, but it has typically meant controlled, engineered, and quantum-coherent manipulation of sound waves at the single-phonon or few-phonon level. Initially, the term appeared in the quantum engineering / quantum information domain: coupling phonons (mechanical vibrations) with superconducting qubits, i.e., wave/phonon control at the quantum level. So far, quantum acoustics has remained wedded to the phonon, i.e., a particle-like picture. In contrast, quantum optics encompasses both photons, which are particles, and waves, i.e., fields, as described in Maxwell's equations. These disparate pictures are unified through the coherent state representation. 

Phonons are particularly suitable for weak fields and control at the single-quantum level. Waves are natural for strong fields, where phonon counting is abandoned in favor of amplitude and phase information. Solid state theory has been almost exclusively carried out in the number state, phonon basis. But the occupation of an audible acoustic mode of 200 Hz at room temperature is $3 \times 10^{10}$ quanta. And for 200,000 Hz ultrasound occupation number is   still  $3 \times 10^7$.  This mode is still at a population $\langle n \rangle  > $ 200,000 at 2 K. Of course there are high energy modes nearly inactive at 2 K, like a 100 cm$^{-1}$ lattice mode.  The coherent state basis is comfortable with such limits, corresponding to barely displaced coherent states. 

\subsubsection*{Why were phonons taboo?}

There have been reasons for suspecting that lattice vibration could not be the root cause of several strange metal (high-$T_c$ material) mysteries, including the persistence of resistivity not only linear in T down to essentially 0 Kelvin, but also at the Planckian ``speed limit''  slope, $1/\tau =k_BT/(m^{*}D).$ This slope, if one is  permitted to use Drude-like theory, corresponds to a Planckian relaxation time $\tau = \hbar/k_B T$. {\it Thermal} phonons acting alone in a flat medium unmarred by disturbances such as  CDW or polarons are  incompetent to cause such scattering  at   very low temperatures. In addition, it has been hard to imagine how the phonons could impose d-wave symmetry on the pseudogaps (we will see how this works through nesting vector influenced CDW, in the coherent state representation, below).  

There were other arguments against a primary role for phonons (or better, lattice vibrations).  The lack of isotope effects seemed to argue that vibrations could not be involved, for example. However, there are problems with this conclusion.  As discussed below, it  may be  incorrect to assume the deformation potential landscape is flat for an electron at low $T$. Lattice effects of the sort that lead to Peierls displacements and Kohn anomalies may be fluctuations, if not permanent features, of the landscape, even at very low T, and   are  examples of strong  electron-lattice  interaction. 

As we shall show in the following,  Planckian diffusion with diffusion constant $D=\hbar/m^*$ is ubiquitous and is the successor to Anderson localization when the random medium is actively evolving. This fact is key to the strange metal regime. A strongly correlated electron-lattice chaotic  ``soup'' {\it emerges}  even from the Fr\"ohlich model, and does not have to be postulated, as it is in the SYK model~\cite{kitaev2015,rosenhaus2019}.

D is independent of any variable except the effective mass. 
This in turn  imposes linear in $T$ resistivity at the Planckian rate.  The resistivity is blind to   any   factor except  as it might change the effective mass of the electrons. Many of the objections to ``phonons'' as the root cause then melt away, such as the lack of isotope effects in experiments, since the Planckian rate remains the same after isotope substitution.

 

Our quantum acoustic realization of the Fr\"ohlich  model (the wave-on-wave or WoW mean field approach, described in section \ref{s:sec2} ) derives its parameters directly from the best experimental estimates of the strange metals,. So far, we have hovered only near optimal doping. Within  WoW, Planckian diffusion, spontaneous polaron formation, charge density waves, pseudogap behavior  with d-wave symmetry, correct Drude peak displacement spectroscopy, and by-pass of the Mott-Ioffe-Regel limit all emerge naturally.   This is making  a strong case for lattice motion coupled to electrons  as the root cause of the strange metal mysteries. Because the question was previously evaluated under a perturbative mindset, lattice vibrations have not been given a fair trial, until now.

\subsection{Tradition}

The coherent state representation of lattice vibrations (leading to waves and fields) has received very little attention compared to the formally equivalent  number state representation.  In fact,   coherent states have been discouraged, in spite of their extreme success and importance right next door in quantum optics, where the classical limit fields  are governed by Maxwell's equations. 

This quote from chapter 24 of the standard and influential text by Ashcroft and Mermin~\cite{ashcroft1976solid}reflects a prevailing attitude:

\begin{quote}
“….we have regarded ….phonons as particles, for which the crucial equations … express the conservation of energy and momentum. However, {\it the same constraints can be derived by viewing the phonons…not as particles, but as waves.} ….This alternative point of view {\it cannot contain any new physics}, but is nevertheless worth keeping in mind for the additional insights it sometimes affords”.[Italics ours].
\end{quote}
The coherent state representation leads directly to the wave picture marginalized just above (see section \ref{coherentstates}). This statement plainly conveys the opinion that serious work requires a particle-based number state, time independent  representation.  Other texts are not so specific but imply as much, through omission. 

One reason the wave picture of lattice motion is unpopular is that it forces one into the time domain, and strongly favors a real space picture of the lattice.  Both are  not traditional. 


We strongly disagree with  ``no new physics'' claim.  With such logic, there is ``no new physics'' in the time-dependent Schr\"odinger equation compared to the time-independent Schr\"odinger equation! That is only true in an extremely  narrow sense.  For most of us, ``new physics’’ will mean the discovery of previously unknown and emergent phenomena not found before. Such discoveries are highly dependent on the starting representation since exact solutions will never be available. 

\begin{figure}[htbp] 
   \centering
\includegraphics[width=0.46\textwidth]{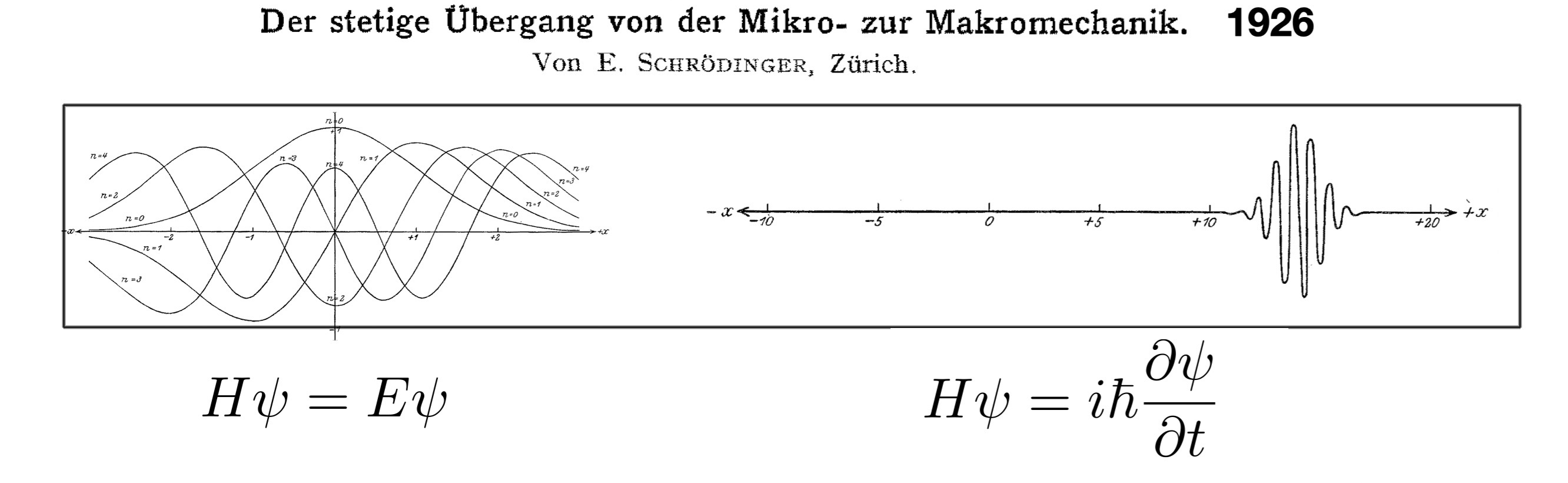} 
   \caption{Photocopy from Schr\"odinger's 1926 paper, showing his drawing of what we now call a coherent state of the harmonic oscillator. In the present context, second quantized number states lie on the left, and coherent states on the right. They do not cancel out each other's ``new physics''. }
   \label{Schroedinger1926.jpg}
\end{figure}

 Schr\"odinger, in 1926, year 1 of quantum mechanics, introduced what we now call the coherent state of an oscillator; see figure \ref{Schroedinger1926.jpg}.
The second-quantized number states live in the energy domain, but the coherent state picture which we advocate here leads inexorably to the time domain. It is not a new model, but rather a perfectly correct representation, one that inspires completely different approaches to classic problems, some unsolved till now.

 \subsection{Coherent states of the lattice}
 \label{coherentstates}
  \begin{figure}[htbp]
     \centering
     \includegraphics[width=0.75\linewidth]{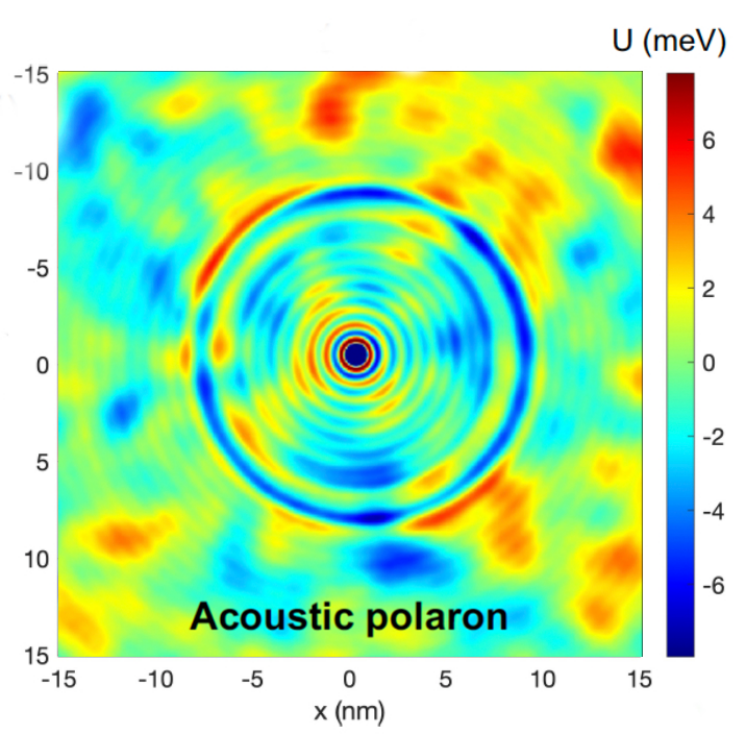}
     \caption{A snapshot of the deformation potential in colorscale after a polaron had recently formed about 3 psec into the WoW propagation. This is a good moment to herald the power of the coherent state representation: The entire scene seen in figure  \ref{polaron3psec}, including the polaron and waves emanating from it, and the thermal  part of the  deformation potential, plus information hidden in the snapshots about the rate of change of all these things, is {\it provided by a single multivariate Gaussian coherent state.} All that complexity is contained in one coherent state configuration.   }
     \label{polaron3psec}
 \end{figure}
 \begin{figure}[htbp] 
   \centering
   \includegraphics[width=3.5in]{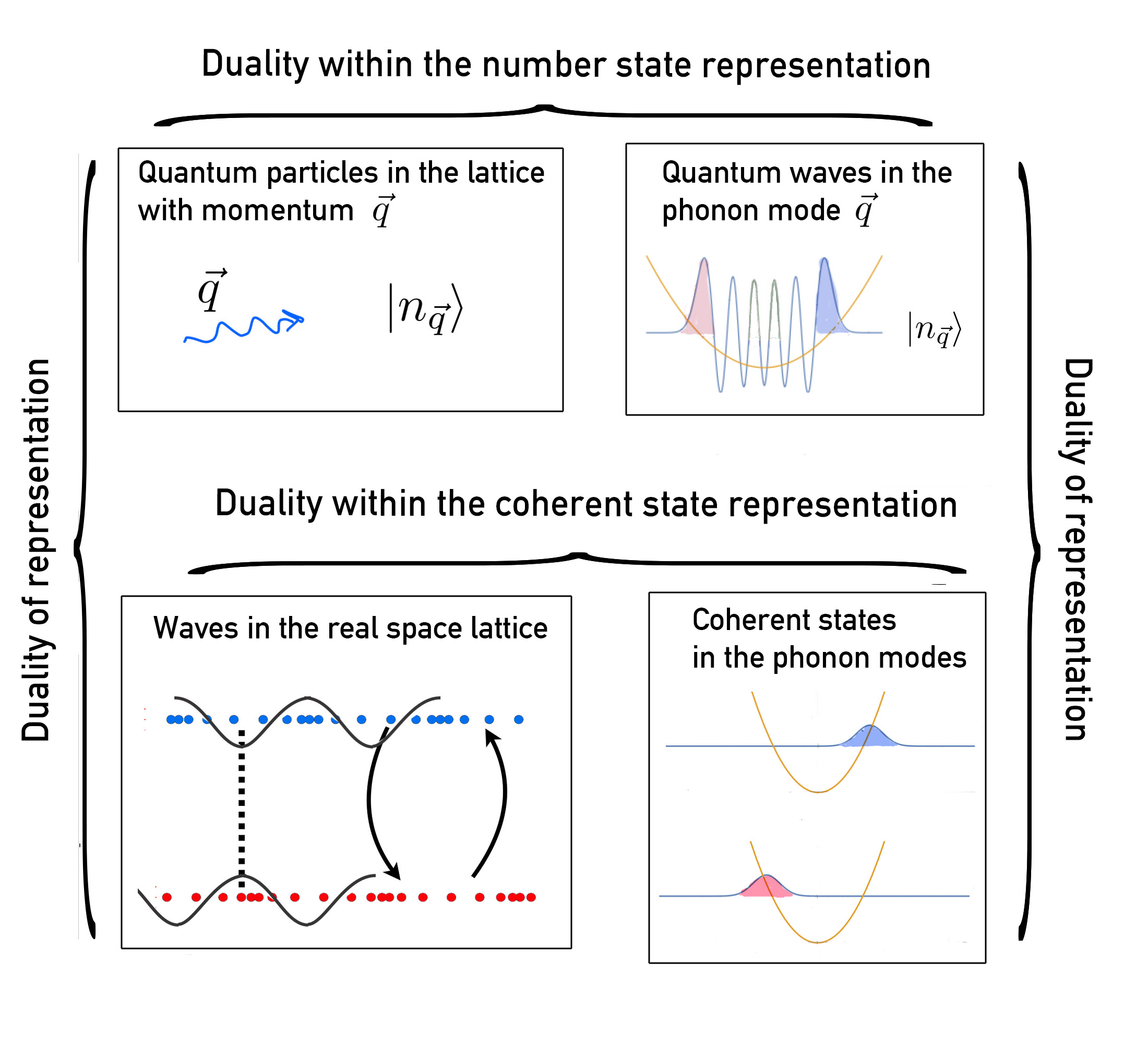} 
\caption{Four wave-particle dualities of crystal lattice quantum mechanics. Reading across in the top row, the Fock $\vert n_{\vec q}\rangle$ state {\bf particle} representation (left) corresponds to a {\bf  wave} in the corresponding $\vec q$ normal mode (right).   In the lower row, a real space vibrational {\bf wave } pattern, (left) lives on the lattice, in which the atoms are all given quite well-defined positions and momenta (within the uncertainty principle). This corresponds to a coherent state {\bf particle}, right, a compact coherent state.    Reading down, two more dualities emerge. In the left column, there is a duality regarding the lattice, with a {\bf particle} representation at the top, and {\bf waves} below.  In the right column, there is a duality regarding the modes, and we go from a {\bf wave} at the top, to wavepacket {\bf particles} below.
 }
  \label{dualdual}
\end{figure}
\begin{figure*}[htbp] 
   \centering
   \includegraphics[width=6.5in]{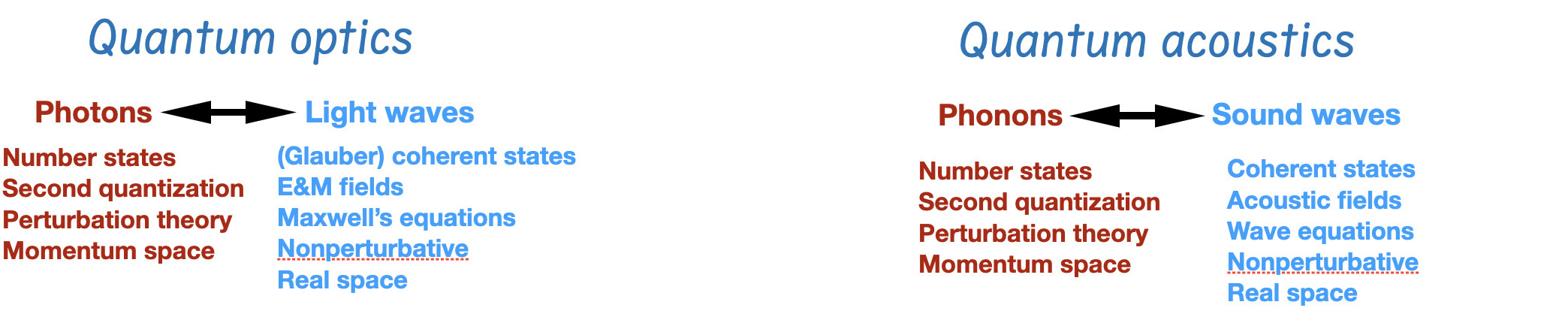}
   \caption{Particles and waves in quantum optics and quantum acoustics. Electrons  interact with the photons, or vacuum light waves on the left, or phonons, or lattice sound waves on the right.}
   \label{opticsacoustics}
\end{figure*}
   The coherent states include the ground state of the oscillator and all possible displacements in coordinate and momentum. Set free to oscillate under the time-dependent Schr\"odinger equation,  the parameters in the wavefunction corresponding to its average position and average momentum follow the corresponding purely classical solutions, as Schr\"odinger proved. Of course, in solid state theory, our oscillators are the normal modes, i.e. the phonon modes, of the lattice. 
 
\subsubsection*{A peek at the polaron} 
 In many body systems, coherent states are remarkably powerful. For example,  the 30 x 30 nm scene in figure~\ref{polaron3psec}, including the polaron, its emitted waves, and the thermal background, is the result of a single multidimensional coherent state in thousands of vibrational degrees of freedom. It is a snapshot of a WoW simulation in progress, where the polaron recently formed spontaneously.





Thermal systems are naturally conceived in terms of coherent states, which are often thought to be what remains after the ravages of decoherence. They are most robust against further decoherence.  
\subsubsection*{Excited number states are cat states}
In contrast, excited number state phonon modes  are actually cat states. The $\vert n\rangle$ phonon mode for $n=10$ seen in at the upper right in figure~\ref{dualdual} is certainly a cat state. The normal mode oscillation is coherently both stretched and compressed, and everywhere in between. It is a prime target for decoherence. At 200 K, a gigahertz mode corresponds to $n\approx 26,000$.  Frankly,  the number state representation is quite uncomfortable here.  One goal in choosing a basis is to come as close as possible to the physical conditions.  Coherent states win hands down over eigenstates in finite-temperature solids.

There is a common misconception that somehow coherent states are intrinsically semiclassical. Not so: they are  equally at home describing motion near the ground state at 1 Kelvin, or much higher, at 1000 Kelvin. They are as pure and complete a quantum basis as are the number states.

Figure \ref{opticsacoustics}  makes clear the perfect analogies of quantum acoustics with quantum optics, and thus gives strong motivation for exploring what the example of quantum optics has to offer.



 We follow the pathway of quantum optics pioneered by Glauber~\cite{glauber1963}, unifying the photon and the electromagnetic field, with the help of the coherent state representation. This becomes the blueprint for the neglected wave perspective of lattice vibrations -- the blueprint for \emph{quantum acoustics}. 
\subsection{The dual duality of waves and particles}
\begin{figure}  [htbp]
   \centering   \includegraphics[width=0.48\textwidth]{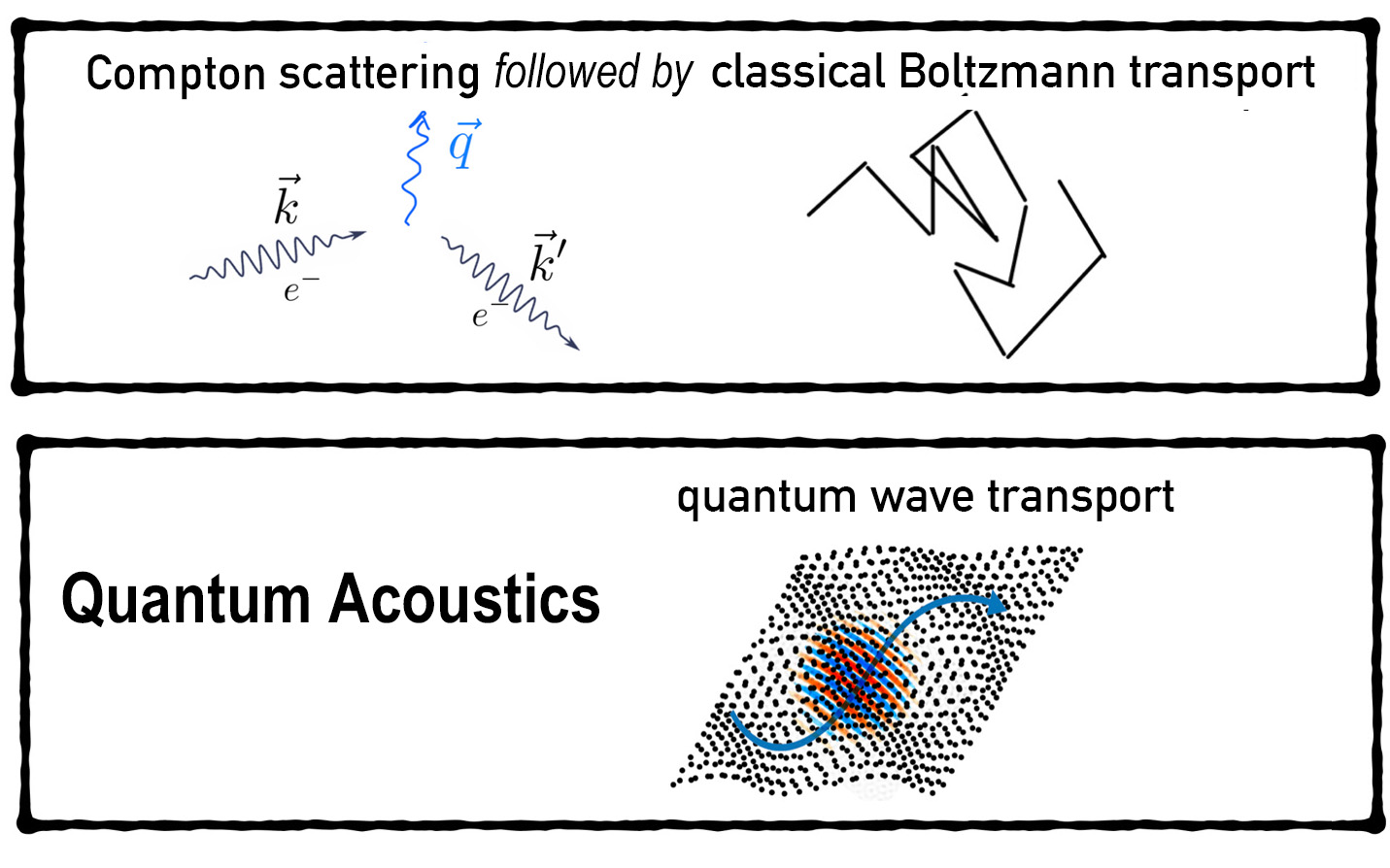} 
   \caption{ At the top, a perturbative Compton scattering picture is characteristic of 70 years of solid state theory, carefully balancing energy and momentum in a local process. Bereft of any easy extension beyond first order, incoherent Boltzmann transport in assumed in order to get to go to higher order in an ersatz way. This is Bloch-Gr\"uneisen theory, which works well for ordinary metals. Below, total energy and momentum, field plus particle,  are not  nearly so carefully balanced, as when an electron deflects in an electromagnetic field for example. Behind the scenes, of course both are conserved. What is shown is a  time dependent quantum electron wave navigating a random thermal, blackbody-like sea of the lattice deformation potential. In our work, the electron and lattice exchange energy and momentum via mean field back-action.  While this neglects quantum entanglements that develop, this is much less of a sin than throwing coherence out altogether beyond first order. }
       \label{usual}
 \end{figure}

   There are close analogies between quantum optics and quantum acoustics. Photons are particles, electromagnetic fields are waves. Phonons are particles, acoustical vibrations are waves.  We hope  figures~\ref{opticsacoustics},  \ref{dualdual}, and \ref{usual}  will make clear the beautiful wave-particle dualities underlying lattice quantum mechanics. But many decades of second quantized perturbation theory presented as inevitable makes it difficult for some to accept that another paradigm could be superior in important circumstances.
   
   It is more than equally valid, it adds deep insight, calculational tools, and numerical results with explanations crucial to the strange metals.  Instead of the energy domain, momentum space, eigenstate,  number state basis, ending with perturbation theory, we are suggesting a real space, time domain picture, in the coherent state basis for the lattice, with the electrons treated nonperturbatively. The former is a particle-like formalistrange metal and the latter a wave-like formalistrange metal. Both lead quickly to insights and results, but they are in completely different arenas. Neither arena should be ignored.

\subsection{The deformation potential
}
\subsubsection*{Bardeen and Shockley}
A common currency of vibronic interactions is the ``deformation potential'', a local change in band structure and band energy due to aggregation and rarefaction of atomic density coming from the dynamical summation of all acoustic phonons present. 
Bardeen and Shockley's introduction of the  deformation potential~\cite{ShockleyBardeen}  marked a turning point in the theory of electron-lattice interactions, and of course electrical properties of materials. It is telling that these authors did not use the term ``electron-phonon'' interactions. The word phonon does appear, but it is clear that Bardeen and Shockley did not think in terms of ``one phonon at a time''. Instead, they emphasized a deformation landscape and its classical evolution, together with quasielastic electron scattering (they used the phrase ``essentially elastic'') within effective mass theory. 
Quoting from that paper, they state 
 \begin{quote}
     This implies that the pertinent acoustical waves can be treated by classical methods, even at fairly low temperatures.
 \end{quote}
Indeed the paper reads as if the deformation potential they introduce is to be used in a semi-classical way. But in  1950, perturbation theory almost had to be introduced, because the semi-classical lattice picture was too numerically demanding. Landau and Pekar used it, but for the specific case of a nonlinear  integro-differential equation which could be solved. It was out of reach to use the equations for more general dynamical evolution. Now, it can be done on a laptop.

The Bardeen-Shockley perturbative approach was subsequently adopted by nearly everyone and has become the standard paradigm. Laying fallow since Landau and Pekar sat a classical field paradigm, also suggested by Bardeen and Shockley. There is one  essential generalization: we must embrace the time dependence of the field to unleash its powers. 

In effect, Bardeen conceived of something very close to the way thermal diffuse scattering (TDS) is done now: the lattice is frozen in a  ``snapshot'' of its motion, caught in the act, so to speak, with typical thermal atomic displacements away from the high symmetry equilibrium positions present.  If the scattering (which is indeed fast and perturbative in TDS using thin films and fast electrons) is computed for this imperfect fixed lattice, a remarkable rendition of the diffuse scattering between the Bragg peaks results. There is no need for averaging over different snapshots  if the sample is large enough. We have not done it, but a time-dependent perturbative treatment of the coherent state representation of the lattice should fully  justify the TDS approach.

The Fock state, Compton scattering, number state perturbation picture  leads to the ``create or destroy a phonon'' rule for pure metal electron deflection. This rule is   problematic, since it implies complete decoherence of the electron by the lattice after any such deflection, because the lattice goes into an orthogonal state.  Boltzmann transport, used to get beyond first order, also  carries   coherence no further than one deflection. The whole approach is blind to coherence beyond one deflection. Treating coherence properly is one of the challenges of working in a many body paradigm of mostly highly excited number states.  It is lucky that the drastic Boltzmann approximation works so well for the normal metals, but its success has in our view delayed understanding of the roiling vibronic sea that is the strange metals.  There, strangely, coherence does matter.  


\subsubsection*{Unfreezing the deformation potential}
 Thus, for 70 + years,    the  paradigm has been  a frozen  deformation potential,   treated  as a perturbation, sandwiched between phonon Fock eigenstates.  Higher order scattering was treated as Boltzmann collisions.  We now are quite certain that this approach  is missing two  ingredients essential to understand the strange metals: (1) time dependence of the deformation potential,  (2) quantum dynamics beyond first order, with coherence of the electron lasting many collision times. 
 
 The widespread perception that  the phonons fail to account for strange metal behavior thus stemmed in part from weaknesses in this traditional Bloch-Gr\"uneisen-Boltzmann  platform. A time-independent, incoherent Boltzmann approach fails to capture the right physics, contributing to the discouraging outlook for ``phonons,'' or better, lattice vibrations, leading to a search for new electron-electron physics. Once the coherent state representation is put into play, new physics emerges that enormously brightens the understanding of the role of the lattice.
 
The coherent state representation shows   the deformation potential to be a  roiling, chaotic, Gaussian sea with very large forces imposed on the electron. 
 The time-dependent local forces due to the thermal deformation potential are large ($10^7$ V/cm) even at 100 Kelvin. We show below that the  thermal deformation potential causes  electrons to diffuse with a Planckian  constant $D = \alpha \hbar/m^*$, with $\alpha$ near 1. When we run our WoW code (section \ref{s:sec2}), for strange metal parameters,  we find that the diffusion is indeed Planckian, over wide parameter ranges.  This leads to linear in T resistivity.  If you care to assign a characteristic time in the strange metal phase, the Planckian diffusion leads to $\tau - \hbar/k_B T$, but it is not clear to what process $\tau$ is supposed to characterize, especially in light of the  vibronic ``soup''~\cite{sun_lorenz_2024} seen in the WoW simulations.


An electron in a  thermal deformation potential is closely analogous to an electron exposed to a very strong blackbody electromagnetic field.  There is a Debye cut-off at high frequencies in the deformation potential, corresponding to a blackbody field as seen beyond a low pass filter. This analogy is extremely useful in guiding and perhaps recalibrating our thinking.

   Using coherent states to describe the lattice brings it alive, giving every atom a nonequilibrium position and momentum limited only by the uncertainty principle, suggesting an active sea of lattice waves.   Once again we call attention to the fact that the whole scene in figure~\ref{polaron3psec} is the result of a single coherent state, a direct product of all the normal mode one dimensional coherent states. Electrons living in this sea are subject to strange metalooth,   time and space-dependent forces, very similar to blackbody fields.

\section{Quantum Acoustics}

\subsection{Prior use of the term ``quantum acoustics''
}   
Before and concurrent with our work, the term ``quantum acoustics'' has not received wide use but has again usually meant a few quanta (phonons)  such as ``Listening For New Physics With Quantum Acoustics,'' about a  phonon-qubit swap device~\cite{linehan_listening_2024}. Or surface acoustic waves~\cite{9880558}. This is a bit different than the parallel context of ``quantum optics'', which although often studied near the single photon limit, is also perfectly comfortable with classical electromagnetic fields.  We are not suggesting a  change in terminology anywhere, except we are broadening the term quantum acoustics to include everything from individual phonons, to classical limit acoustical waves. In this way it comports with quantum optics.




\subsection{Wave-on-wave method}\label{s:sec2}

The recently developed wave-on-wave or WoW method~\cite{Aydin24} arises almost as a necessity if one adopts the wavelike, time dependent, coherent state   representation of the  lattice.   It empowers nonperturbative and coherent  treatment of  electron-lattice vibration interaction. Forces of the lattice act on the electron, and the electron back-acts on the lattice. We derive this now starting from a Fr\"ohlich Hamiltonian.

We describe an electron by a quantum wavepacket and the lattice vibrations  by coherent states, not number states. (We eschew the word phonon on our context, because it refers to a quantized particle.  We treat the lattice along the lines of the classical electromagnetic field  limit of quantum optics). Employing the deformation potential approach in this way and in real space, the electron is  confronted   with a disordered and dynamic landscape attributable  to thermal lattice vibrations. Electrons quasi-elastically deflect from the formidable bumps and hills of this ever-changing potential. 
\subsubsection*{Fr\"ohlich Hamiltonian}
More than 70 years later, we have extended the program begun by Bardeen and Shockley~\cite{hellerkim}, and  Fr{\"o}hlich, whose Hamiltonian is seen second-quantized form~\cite{Many-Particle}, equation~\ref{froh}.  It was always known it contained the potential for correctly describing very strong electron-lattice interactions, since Fr\"olich's first papers introduced and derived polarons with it~\cite{frohlich1954electrons}.
\begin{eqnarray}
\begin{aligned}
\hat{H} &=\hat{H}_{\text{e}}+\hat{H}_{\text{ph}}+\hat{H}_{\text{e-ph}} \\
&=\sum_\mathbf{k}\epsilon_\mathbf{k} c_\mathbf{k}^{\dagger} c_\mathbf{k} + \sum_\mathbf{q} \hbar\omega_\mathbf{q}(a_\mathbf{q}^{\dagger}a_\mathbf{q}) \\
&+ \sum_{\mathbf{k},\mathbf{q}}g_{\mathbf{k},\mathbf{q}}c_{\mathbf{k}+\mathbf{q}}^{\dagger} c_\mathbf{k}(a_\mathbf{q}+a^{\dagger}_{-\mathbf{q}}),
\end{aligned}
\label{froh}
\end{eqnarray}
where $\epsilon_\mathbf{k}$ is electron band energy with $\mathbf{k}$ wavenumber, $c_\mathbf{k}^{\dagger}$ ($c_\mathbf{k}$) is creation (annihilation) operators for electrons, $\omega_\mathbf{q}$ is the frequency of the phonon normal mode $\mathbf{q}$, $a_\mathbf{q}^{\dagger}$ ($a_\mathbf{q}$) is creation (annihilation) operators for phonons, and $g_{\mathbf{k},\mathbf{q}}$ is electron-phonon coupling strength. In the Hamiltonian, we have omitted the zero-point energy of the lattice, as it constitutes a constant offset that does not influence the dynamics or interaction terms relevant to the analysis of polaron formation.

Following the path forged by quantum optics~\cite{scully1997quantum, walls2007quantum}, we describe the evolution of the deformation potential using coherent states $\vert \alpha_{\mathbf{q}} \rangle$, characterizing the dynamics through their expectation values $\alpha_{\mathbf{q}} = \langle \alpha_{\mathbf{q}} \vert a_{\mathbf{q}} \vert \alpha_{\mathbf{q}} \rangle$, rather than evolving the phonon field operators directly. We initialize each mode in a thermal coherent state,
\begin{align}
\alpha_\mathbf{q}(t_0)=\sqrt{\langle n_{\mathbf{q}}\rangle_{\text{th}}}e^{i\phi_\mathbf{q}},
\label{alphaint}
\end{align}
where $\phi_{\mathbf{q}}$ is a random phase and the thermal amplitude is given by $\langle n_{\mathbf{q}}\rangle_{\text{th}} = \left[\exp\left(\hbar\omega_{\mathbf{q}}/k_{\textrm{B}} T\right) - 1\right]^{-1}$.


At thermal equilibrium, the modes are  treated as having been in contact with a heat bath at temperature $T$.  Because there are infinitely many normal modes in any  cone in $\bf k$ space, the statistical ensemble is boiled down to simple average population. By taking into account the independence of normal modes, the collective lattice vibration $\vert \chi \rangle$ can be described as the product of coherent states of the individual normal modes -- essentially a multimode coherent state $\vert \chi \rangle = \otimes_{\mathbf{k}} \vert \alpha_{\mathbf{k}} \rangle$, as  discussed in Ref.~\cite{hellerkim}. 

\subsubsection*{Deformation potential}
The quantum field of the deformation potential is written as the gradient of the displacement field $\hat{\mathbf{u}}(\mathbf{r},t)$
\begin{eqnarray}
\begin{aligned}
    \hat{V}_D(\mathbf{r},t)
    &=
    E_d\nabla\cdot\hat{\mathbf{u}}(\mathbf{r},t)
    \\
    &=
    -\sum_{\mathbf{q}}
    g_{\mathbf{q}}
    (a_{\mathbf{q}}e^{-i\omega_{\mathbf{q}}t}
    +
    a_{-\mathbf{q}}^\dag e^{i\omega_{\mathbf{q}}t})
    e^{i\mathbf{q}\cdot\mathbf{r}},
    \label{VDquantumfield}
\end{aligned}
\end{eqnarray}
By taking the lattice modes to be thermally (but quantum mechanically) populated, the Hamiltonian of Eq.~\ref{froh}  gives rise to a quasi-classical, dynamic lattice deformation potential
\begin{eqnarray}
\begin{aligned}
    V_D(\mathbf{r},t) &= \langle \chi \vert  \sum_{\mathbf{q}}
    g_{\mathbf{q}}
    (a_{\mathbf{q}}e^{-i\omega_{\mathbf{q}}t}
    +
    a_{-\mathbf{q}}^\dag e^{i\omega_{\mathbf{q}}t})
    e^{i\mathbf{q}\cdot\mathbf{r}} \vert \chi \rangle \\ &=
    \sum_{\mathbf{q}}^{\vert \mathbf{q} \vert \le q_D}
    g_{\mathbf{q}}
    \sqrt{\langle n_{\mathbf{q}}\rangle_{\textrm{th}}}
    \cos(\mathbf{q}\cdot\mathbf{r}-\omega_{\mathbf{q}}t+\varphi_{\mathbf{q}}),
\end{aligned}
\label{deforma}
\end{eqnarray}    
where $q_D (\omega_{\mathbf{q}})$ is Debye wavenumber (frequency), $\mathbf{r}$ is continuous position, $\varphi_{\mathbf{q}}=\textrm{arg}(\alpha_{\mathbf{q}})$ is the (random) phase associated with a coherent state and the mode population is determined by the Bose-Einstein occupation $\langle n_{\mathbf{q}}\rangle_{\textrm{th}} = [\exp (\hbar\omega_{\mathbf{q}}/k_{\textrm{B}} T) - 1]^{-1}$. 

Equation~\ref{deforma} defines the {\it thermal} part of the deformation potential.  There is an addition to be made, coming from back-action of the electrons on the lattice.

  Within the present deformation potential framework, an electron undergoes continuous quasielastic, coherence-preserving deflection while roaming through the slowly altering (but Anderson localization destroying) potential landscape. The hills and valleys act like a mobile defect field, even in a perfect crystal, with the key twist that the defects are constantly morphing from one shape to another. 

The deformation potential enters into the Hamiltonian of an electron as an internal potential field that is treatable nonperturbatively:
$\mathcal{H}_{\textrm{el}}(\mathbf{r}, \mathbf{k}, t)  = E_0(\mathbf{k}) + V_D(\mathbf{r}, t)$ 
where $\mathbf{k}$ is the electron wavevector.  The band energy of the underlying, undistorted lattice, $E_0$, can be described within widely used effective mass model. Alternatively it can be extracted from density functional theory or experiments. We simultaneously solve the wavepacket dynamics utilizing the split-operator method, and the corresponding wave equation for the
 time-dependent deformation potential. The employment of (Gaussian) wavepackets provide a versatile, and insightful, framework to solve different problems in condensed matter physics. We call this the wave-on-wave, or WoW approach. Subsequently, the WoW approach enables us to deduce electronic transport properties either within  the diffusion picture~\cite{heller22} or within the Kubo formalistrange metal. 

 \subsubsection*{Mean field and back-action}
Motivated by Refs.~\cite{leopold2022landau} and~\cite{leopold2021derivation}, we consider the evolution of an initial state $\vert \Psi \rangle$ of product form
\begin{equation}\label{Eq:LP_ansatz}
    \vert \Psi \rangle  = \vert \psi \rangle \otimes \vert \chi \rangle
\end{equation}
composed of the electronic state $\vert \psi \rangle$ and the lattice state $\vert \chi \rangle$. As a variational ansatz, the product state  Eq.~\ref{Eq:LP_ansatz} obeys the Landau–Pekar-like equations ~\cite{pekar1946local, landau1948effective}, the coupled, nonlinear equations of motion,
\begin{subequations}
\label{coupled}
\begin{align}
    i\hbar \frac{\partial \psi}{\partial t} & = \left[\frac{1}{2m} (i\hbar\nabla + e\mathbf{A})^2  + e\varphi +  V_D(\mathbf{r},t) \right]\psi, \\
    i\hbar\frac{\partial\alpha_\mathbf{q}(t)}{\partial t} & = \hbar\omega_\mathbf{q}\alpha_\mathbf{q}(t) + g_\mathbf{q} \int e^{-i\mathbf{q}\cdot\mathbf{r}}|\psi(\mathbf{r}, t)|^2\, \textrm{d}\mathbf{r}.
\end{align}
\end{subequations}
along with the vector $\mathbf{A}$ and scalar potential $\varphi$ stemming from an static, external magnetic $\mathbf{B} = \nabla \times \mathbf{A}$ and electric field $\mathbf{E} = -\nabla\varphi$. 


Construction of the deformation potential   equation according to equation \ref{deforma} still holds, but the single mode coherent states no longer evolve freely starting from the initial condition equation \ref{alphaint}, and instead are affected by the mean field back action term in equation \ref{coupled}.



In the second of equations \ref{coupled} we see the force that an electron exerts on the coherent state $\alpha_\mathbf{q}$ of the normal mode $\mathbf{q}$ explicitly. The coherent state of  mode $\mathbf{q}$ interacts strongly if $|\psi|^2$ has strong components at wavevector $\mathbf{q}$.  Therein lies a lot of physics, quite rightly new physics, arising from a direct time-dependent realization of the Fr\"olich Hamiltonian, including charge density waves and polaron formation. This approach, and the universality of Planckian diffusion discussed below, provide   strong  and  direct links to  various strange metal behaviors.

\subsubsection*{Landau and Pekar}    
Remarkably, Landau and Pekar~\cite{landau1948effective,pekar1946local} arrived at the quantum acoustic equations with back-action in the form of a polaron theory  in 1948, as made clearer and rigorous by Frank {\it et. al.}~\cite{frank_zhou_2017} This was of course before the introduction of Bardeen's deformation potential, and before second quantized perturbation theory became popular. Frank {\it et. al.} showed that the Fr\"ohlich Hamiltonian leads to 
Landau and Pekar.

\begin{figure}[htbp]
\centering
\includegraphics[width=0.45\textwidth]{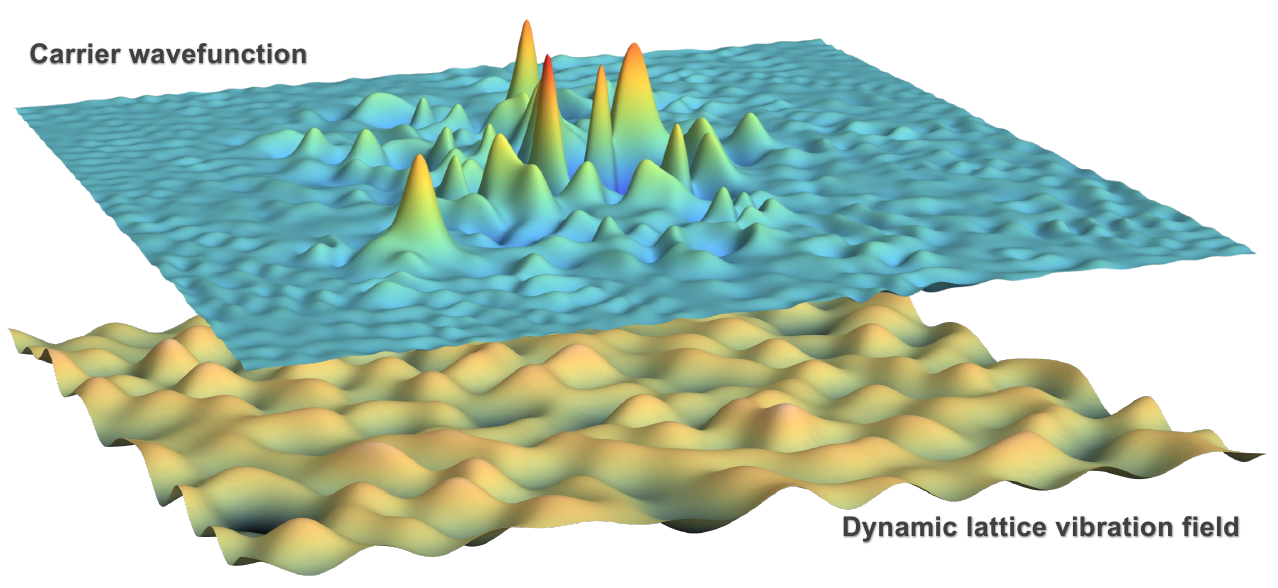}
\caption{Wave-on-Wave approach. Nonperturbative and coherent dynamics of charge carriers and thermal lattice vibrations. A snapshot of a carrier wave packet (the magnitude  is shown on top) coherently propagating under a spatially continuous, dynamic disorder field (bottom) formed by acoustic lattice deformations. A charge carrier quasielastically scatters (similar to impurity scattering) in this disordered landscape. }
\label{coherent}
\end{figure}
\subsubsection*{The remarkable properties of the deformation potential}
The deformation potential $V_D,$ equation~\ref{deforma}, is statistically the same if inverted. It averages to $0$, over infinite areas or along any infinite line.  It also averages to $0$ over time at any one spot.  (This is important for transport, since the electrons can't get too comfortable in any one place. Basins will turn to mountains). The potential is random but uniformly so, and has the form of a blackbody radiation field~\cite{HBT1}. 


The quantum acoustics approach gives rise to an undulating, time-dependent potential landscape, as found in Ref.~\cite{heller22}.  The relationship to quantum transport is found there and in Ref.~\cite{PNASpr}: 


     The coupling constant $E_d$ (equation \ref {VDquantumfield}) is usually given in electron volts.  It is defined as 
     \begin{equation}
     \label{Ed}
         E_d = V\frac{dE}{dV}, 
     \end{equation}
     the change in local Fermi energy as the volume changes. This can be quite large, on the order of 10 to 30 or more electron volts. 

This has a remarkable implication, even for a modest $E_d$ of 10 ev. Assume that the volume fluctuation of a crystal at 100 K is a part in $10^4$, which is also modest. For such a fluctuation over the space of a nanometer, electrons are therefore experiencing random blackbody field gradients on the order of millions of volts per meter, corresponding to a blackbody radiation field at around 5000 Kelvin.   An impressive number  at 100 Kelvin!  This is just one indication of many that the deformation potential should be taken more seriously than is possible in first-order perturbation theory, at least for the strange metals where the deformation potential looms much larger on the scale of the Fermi energy than in the ordinary metals.

\subsection{Path-integral road to quantum acoustics}
\label{path}

 To investigate the accuracy of the mean field approximation in the quantum acoustics framework, the full system dynamics can be simulated. This is made possible by an influence functional approach~\cite{FeynmanInfluence}. We can use the method to get close to exact results. Since the deformation potential, as given in equation~\ref{deforma}, is linear in the lattice coordinates, and the initial state of the lattice is a coherent Gaussian  state,  the influence functional can be calculated exactly. The   functional can be decomposed into the mean field approximation, and the full system dynamics corrections. These corrections can be ``unraveled'' into Gaussian noise~\cite{STOCKBURGER2001249} and a stochastic equation for the reduced density matrix can be derived. This procedure is formally exact, and can be used to verify the accuracy of the mean field approximation. 

 We have developed such  a path-integral approach to quantum acoustics, providing a pathway to exact treatment of electron mobility in the presence of the lattice (for  linear coupling such as in the Fr\"olich model), and reaffirming the basis of our mean field approach\cite{joost}.
Our preliminary   results doing just this are very encouraging~\cite{joost}, suggesting that the strange metal phenomenology is in the ``safe'' strong coupling limit of mean field approximation.
 Within the coherent state picture, we formulated a non-Markovian, stochastic master equation that captures the exact dynamics of any system linearly coupled to a harmonic lattice. We applied the formulation to procedure to the venerable Fr{\"o}hlich model.
 

Even though  we trace  the bath away, we are still able to recover the expectation values $\hat O(t)$ of certain observables via the Ehrenfest theorem.   These can  be evaluated as 
\begin{equation}\label{eq:expectation_value}
    \langle\hat O \rangle (t) =\left<\langle\psi_-(t)|\hat O|\psi_+(t) \rangle\right>_W.
\end{equation} 
For instance, the expectation values of the position and momentum operators for a lattice mode $\boldsymbol{q}$ are given by
\begin{align}
\label{eq: Ehrenfest}
    \frac{d}{dt} \boldsymbol{X}_{\boldsymbol{q}}(t) = \sigma_2\cdot\left(\omega_{\boldsymbol{q}}  \boldsymbol{X}_{\boldsymbol{q}}(t)  +\frac{1}{2}\langle \boldsymbol{g}_{\bm q} \rangle(t) \right)
\end{align}
where $\sigma_2$ is the Pauli matrix and the expectation values of $\bm g_{\bm q}(\bm r)$ are determined by Eq.~\ref{eq:expectation_value}. Using the Ehrenfest theorem leads to coupled linear differential equations mimicking the classical motion of a bath composed of harmonic oscillators.
The approach has advantages over both the more familiar Lindbladian alternatives~\cite{weiss2012quantum}  and the  Lee-Low-Pines transformation~\cite{PLLTransformation}, which is natural at 0K but not convenient for thermal coherent states.
A summary of the derivation is given in the Appendix.

\subsection{Quantum acoustics applied to normal metals}

In reference~\cite{heller22}, it was shown that the number state perturbation-Boltzmann  theory and coherent state nonperturbative pictures give the same results, lending credence to both.  That is a reassuring platform from which to jump off into the  world of the strange metals, which we now can do within quantum acoustics and WoW.

The initial Gaussian wave packet describing a single electron with an initial average (canonical) momentum $\hbar k_F$ to the $\hat{x}$ direction is given by
\begin{eqnarray*}
\begin{aligned}
    \psi(x,y,t_0)=\frac{1}{2\pi\sigma_x\sigma_y}e^{-\frac{(x-x_0)^2}{2\sigma_x^2}
    -\frac{(y-y_0)^2}{2\sigma_y^2}+ik_Fx},
\end{aligned}
\end{eqnarray*}
which lives in the deformation potential $V_D(\mathbf{r},t)$, with or without an external magnetic field. It becomes an unruly 2D wavefunction as it diffuses in the deformation potential using split operator FFT in the WoW code. 

 Instead of perturbation theory and incoherent Boltzmann transport approximations, as in Bloch-Gr\"uneisen theory, we nonperturbatively solve the time-dependent equations for the lattice coherent state waves interacting with the Sch\"odinger electron wave, using mean field back-action~\cite{heller22}. Agreement is excellent with the low and high-temperature 2D and 3D Bloch-Gr\"uneisen approach in the weak coupling limit for normal metals: resistivity rises as T$^5$ (Bloch's T$^5$ law) in a pure metal at low T, rolling over to T as the Debye temperature is approached from below (the rollover can start at much lower temperature in some electron poor metals). It goes as   T$^4$ rolling over to T in 2D metals. Specific prefactors are known.  These results were long ago derived by the number state, Compton-Boltzmann perturbation theory.  A more accurate, nonperturbative theory must agree with earlier methods when perturbation theory applies, and quantum acoustics does agree.

 \subsection{``Phonons'':  in or out?}

We have already said that arguing over phonons is misplaced; the term ``phonon'' evokes perturbation theory, which is entirely inadequate to understand the strange metals. Lattice modes can be occupied by millions of phonons; one should instead work in the coherent state representation.

We argue that lattice distorions, if not fixed CDW and polarons, may be part of the landscape at very low T.  These are  poorly or only loosely characterized as phonon effects.  Rather they are nonperturbative lattice mode effects.

 \begin{figure}[htbp]
\centering
\includegraphics[width=0.45\textwidth]{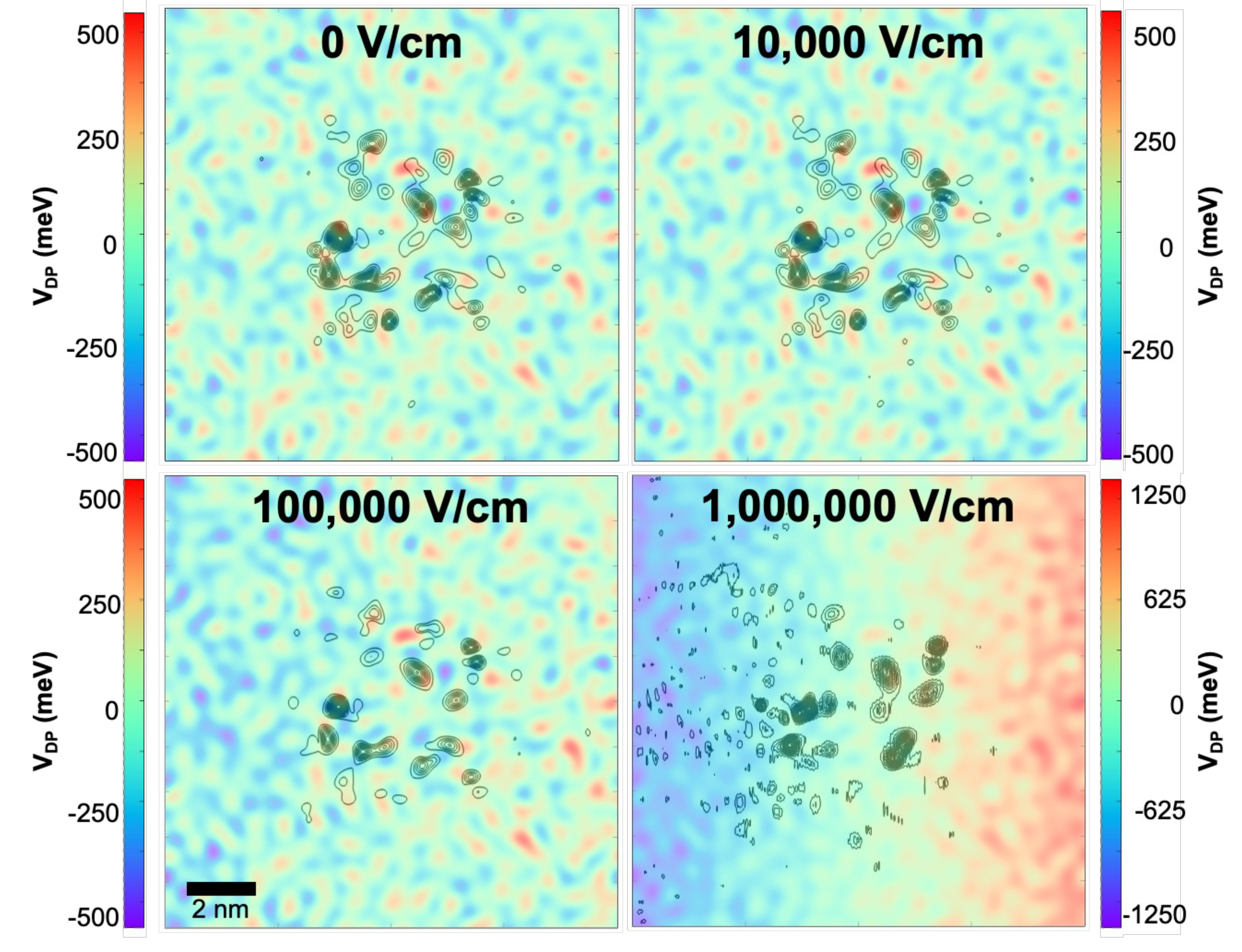}
    \caption{Absence of drift mobility in the strange metal phase. Snapshots of four WoW runs with different applied electric fields, including back-action, are shown in an evolving deformation potential $V_{DP}$ seen in color.   After 0.5 ps of time evolution, starting with a localized wavepacket, the electron density (black contours) is unaffected  in all but the last case.  The lattice was at 120 Kelvin, with  deformation potential constant $E_d = 10$ eV. The frames are 20 nm in each dimension. Apart from backaction changes, the evolution of the deformation potential was the same in each case.  At $10^5$ V/cm, or $0.2$ eV across the window,  the applied field becomes  visible.  The deformation potential  supports local gradients exceeding $2 \times 10^8$ V/m.  The resulting transport must be entirely diffusive, in the sense of Fick's law and population gradients}
    \label{nomobility}
    \end{figure}

 \section{Two key findings with WoW}
 Empowered with the quantum acoustical, WoW approach, some key facts emerged concerning electrons interacting strongly with the lattice, coherently and nonperturbatively. 


\vskip .1in
(1) {\bf Diffusive transport prevails}
\vskip .1in
The key property, and the source of a lack of quasiparticles, Fermi surfaces, etc., is seen in figure~\ref{nomobility} (A. Graf et. al, unpublished).. This is a snapshot of a WoW simulation after 0.5 ps of diffusion, with varying strengths of external fields applied. It demonstrates quickly that transport is {\it  diffusive}, not drift. Strange metal phase transport takes place through chemical potential gradients, as in some semiconductor regimes. The four panels of a WoW simulation show that the electron (black contours) is nearly immune to even huge external fields (100,000 V/cm) due to the mountainous deformation potential.  

Semiconductors, which carry anaolgies to the low doped and optimallly doped  cuprates, are understood to transport by both diffusion and drift, as embodied in the drift-diffusion form 
\begin{equation}
     \mathbf{j} = -D\nabla n + \mu n \mathbf{E}
     \label{driftdiff}
\end{equation}
The semiconductor drift–diffusion equation  is the macroscopic realization of the fluctuation–dissipation relation derived by Kubo\cite{kubo2012statistical}. Both terms can be present and either term can dominate. The strange metal - pseudogap-quasiparticle phases of doping and temperature embody this crossover, with diffusion dominating the strange metal phase. 

The conclusion of the Ramshaw group {\it et. al.}\cite{grissonnanche_linear-temperature_2021} is that an anisotropic scattering rate that has a d-wave–like form and is  consistent with Planckian dissipation at all points on the Fermi surface.
Strange-metal behavior doesn’t require “hot spots” as distinct microscopic regions — rather, the scattering anisotropy is a continuous function around the Fermi surface. They see an isotropic Planckian component and an angle dependent scattering rate.

\vskip .1in
(2) {\bf The Diffusion is Planckian}
\vskip .1in
The diffusive component is Planckian,    $D\approx \hbar/m^*$. We reinforce  this claim below.

In the strange metal phase, the electrons find themselves in a badly distorted landscape, necessisarily becoming strongly correlated with the lattice. They are scraping bottom so to speak, in ways that they are protected from doing in copper for example. Since the electrons strongly interact with the same lattice, they can become strongly correlated with each other. 

These findings comport with and enhance Hartnoll's 2015 conjectures\cite{Hartnoll2015}, and contributions by Nussinov and Chakrabarty\cite{Nussinov_ann.phys_443_168970_2022, NUSSINOV2022168970} regarding diffusive transport at the Planckian rate $\tau = \hbar/k_B T$  in strange metals. The engine is indeed diffusion, but the fundamental element is Planckian diffusion $D = \hbar/m^*$ independent of temperature, see next section \ref{linearT}.

It is clear that the Drude picture does not apply to the strange metal phase; an  electric field short of millions of volts per cm has no visible effect on the  electron motion, see figure \ref{nomobility}. Such fields can be read off directly from that figure. Transport is purely diffusive,  driven by concentration gradients. From one point of view, the strange metals at optimal doping are like doped semiconductors, with a large band gap.






Our WoW runs show this behavior dramatically\cite{Aydin24}:  the diffusion $\langle r^2\rangle_t$ increases linearly at the same rate, substantially independent of temperature, electron-lattice coupling, sound speed, or anything else, except as it might affect the effective mass of the carrier.




This diffusive transport is the new physics of the strange metal phase, and   is indeed  the sought after substitute for the traditional quasiparticle dynamics. Let us be clear: the diffusive transport is entirely attributable to electron-vibration interaction; e-e  plays no role.  If SYK has  relevance, its {\it ad hoc} assumption of chaos has now been given a palpable, microscopic origin - entanglement with vibrations, indeed bosons.


    Ultimately, the driving force is an electrochemical potential gradient. Both terms in the drift-diffusion equation \ref{driftdiff} contribute to $D$, and in some ways the distinction is  not important.  However, physically, what we see in the strange metal simulations is that the transport  is pure diffusion current. 
\section{Linear  in T resistivity from Plankian diffusion }
\label{linearT}

By calling upon charge conservation, Fick's law of diffusion, Ohm's law, and relating the population gradients to chemical potential gradients {\it via}  compressibility $\chi$:  \( \nabla n = \chi\,\nabla\mu \).
      we find the relations:
      \[  \sigma\,\mathbf E
          = -D\,\chi\,\nabla\mu
          = e\,D\,\chi\,\mathbf E
          \quad\Longrightarrow\quad
          \sigma = e^{2}\chi D.
      \]
 This is an Einstein relation, almost self-evident in its physical content.
 Diffusion current stems from Fick's law population filling, and $ \chi$ records how   a chemical potential drop due to an applied field changes the populations. 

Since  we have it on good authority (WoW calculations) that $D$ is constant with temperature, and experimentally $\sigma \propto 1/T$, the conclusion must be that, one way or another, $\chi\propto 1/T$. This is the classical, nondegenerate limit of compressibility for electrons. 
 \begin{figure}[htbp]
\centering
\includegraphics[width=0.45\textwidth]{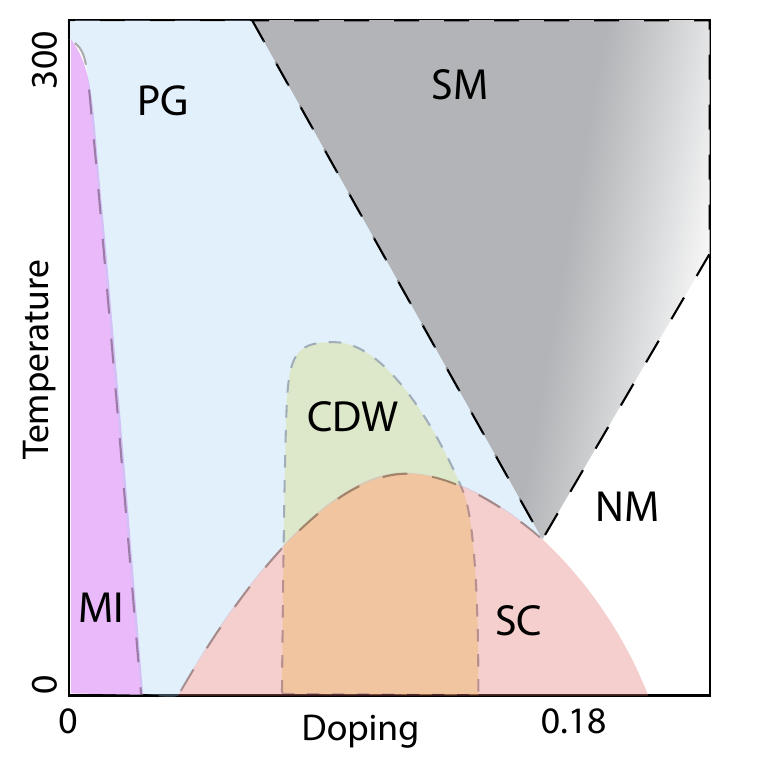}
    \caption{Schematic diagram of  phases of typical cuprates as a function of temperature and hole doping, showing strange metal, normal metal, superconducting, charge density wave, pseudogap, and Mott insulator regions.  Only the superconducting phase boundary is perfectly sharp. The strange metal zone possesses linear in T resistivity, with no typical Fermi surface,  Planckian diffusion, and nondegenerate carriers. The PG zone shows partial return of  partial return of Fermi surfaces and degeneracy. For almost 40 years,  every phase and their boundaries  except perhaps the Mott phase   has been enigmatic.}
    \label{phasedia}
    \end{figure}

We have an idea how this classical behavior arises, and furthermore why the strange metal-pseudogap  boundary exists and looks the way it does (figure \ref{phasedia}). The strange metal phase gives way to the PG phase along the sloping line shown. 

Holes appear near the top of the valence band. Assuming a large gap to the next band, if the potential landscape were flat and steady, the holes  would form a Fermi distribution, only cut off by the termination of the band. At optimal hole doping around 0.18, the chemical potential has been lowered perhaps 40 meV and degeneracy would set in, if the potential landscape were flat and steady. 

It is likely neither flat nor steady. The Fermi distribution is expected to be modulated by both potential energy features and nonadiabatic dynamics of the potential.  We   don't   have a full grasp of this complex, many body regime, but there seems to be  evidence that the electrons effectively become nondegenerate in the strange metal regime.

It may be easier to approach effective nondegeneracy as more holes are provided by doping, and as temperature rises, both increasing the possible room for unruliness of the electrons. Meanwhile in the bulk of the distribution   a  Fermi sea is maintained, so the holes can be  nondegenerate without a collapse of the sea.



`

   \section{  Planckian diffusion: The ghost of Anderson Localization}
      \label{ghostsec}

      Here we give a glimpse of \cite{ghost}, citing  numerical calculations using strange metal parameters,  an experiment involving electron mobility on solid hydrogen,   numerical calculations under more general cases of active media, with and without a temperature, and  a solvable real-space quantum diffusion model.  Each of these highlights different aspects of Planckian diffusion and reinforces  its ubiquity.
      
\label{Planckdiff}
In our paper, ``Planckian Diffusion: The ghost of Anderson Localization''~\cite{ghost}, we demonstrate that once Anderson localization is destroyed by the real-time motion of a random field, an itinerant ``ghost'' emerges in its place: Planckian diffusion, characterized by $D \approx \hbar/m^*$. For this to apply, the medium must move quickly enough for the carrier to be transported non-adiabatically, an easily satisfied condition in most random media. In short, Planckian diffusion  governs the strange metal phase: it is the leading cause and key symptom of  strange metal behavior.
This discovery  has   wide implications, within and beyond the strange metals. To re-state: Quantum particles living in a  random  medium undergoing dynamic change, in a medium that would Anderson localize the particles if it were stationary, will   instead result in their diffusion with a  constant  $D\sim \hbar/m^*,$ independent of temperature, independent of the coupling strength to the lattice.   There will be  little  dependence on the isotopes present, or rate of mixing of the medium, or even the localization length in the absence of  disturbances. This  trend toward a kind of universality neutralizes many of the objections raised over the possible role of the lattice modes in strange metal behavior.  The only system specific dependence is to the effective mass. 

This Planckian diffusion is appropriately called  a ghost\cite{ghost}, generalizing   Anderson localization to moving media. It is more ubiquitous than    Anderson localization, in the sense that  the multitude of  the  localization scenarios is multiplied by the multitude of the ways of breaking them by activity of the medium.  A huge variety of scenarios thus lead to approximately the same   quantum diffusion constant.   


There can be exceptions: If the medium evolution is extremely slow, adiabatic rules would dictate a nonuniversal diffusion constant which must necessarily depend on the rate of change of the potential.  It would not be Planckian. But this rate turns out to be so slow for the case of electron diffusion that we will ignore it for the purposes here. A vast, nonadiabatic  and nearly universal regime lies just beyond ultra-slow. 



In the strange metal regime of the high $T_c$ superconductors,  we have found numerically that diffusion of an electron wavepacket in the active deformation potential field lives within a quantum Planckian  upper and lower bound, $D =\alpha  \hbar/m^*$, where $m^*$ is the effective mass, and $1/2\le \alpha\le 2, $ independent of temperature or electron-phonon coupling strength~\cite{PNASpr}. Using our WoW code, we watch an electron diffuse at almost the same rate after the temperature and coupling changes by orders of magnitude.

  As with the related Planckian speed limit $\tau \sim \hbar/k_B T$,  we call the diffusion Planckian if $D$  is within a factor of 2 of $\hbar/m^*$. These upper and lower limits can be violated, so we call it a ``ubiquitous" phenomenon.

The closely related Planckian speed limit  has received much more attention. See, e.g., Refs.~\cite{Zaanen_Sci.post_6_061_2019, Hartnoll_rev.mod.phys_94_041002_2022, Mousatov_nat.phys_16_579_2020, Zhang_pnas_116_19869_2019, Lucas_phys.rev.lett_122_216601_2019, Gleis_phys.rev.lett_134_106501_2025}). This idea rests in  a compelling but somewhat \emph{ad hoc} way on the time-energy uncertainty relation, $\Delta t \Delta E \sim \hbar$, assuming the characteristic energy scale $\Delta E$ to be associated with thermal $k_BT$. Although exactly what physical process suffers the corresponding time scale $\Delta t \sim \tau$ is sometimes vague, it can be connected to momentum relaxation~\cite{Hartnoll_rev.mod.phys_94_041002_2022}. Nussinov and Chakrabarty have   provided a more concise perspective\cite{Nussinov_ann.phys_443_168970_2022,NUSSINOV2022168970}.

 Planckian diffusion $D=\hbar/m^*$ encompasses  the Planckian speed limit  $\tau \sim \hbar/k_B T$. It is more general in the sense that it applies to systems not possessing a temperature or any equilibrium, and it doesn't require a model for what $\tau$ represents.  The speed limit form for thermal systems is reached from $D=\hbar/m^*$  by an Einstein relation.



  \subsection{Strange Metals}
The original hint about the ubiquity of Planckian diffusion was  seen over a range of strange metals and temperatures, in runs of  the WoW code at optimal strange metal doping parameters and various temperatures\cite{Aydin24}. The random medium is the thermal deformation potential, which is in motion because   its component acoustic waves are moving at the sound speed.  Snapshots of the potential at random times look identical statistically, but they differ everywhere in detail, with the earliest differences in time coming in at the shortest wavelength and highest frequency components of the deformation potential. 

A remarkable lesson about   ghost Anderson Planckian diffusion is learned:  in its universal regime,   if it is Planckian for one set of parameters, it stays Planckian with nearly the same diffusion   constant even if shorter or stronger oscillations are added to the random potential. Or there could be fixed impurities added to the mix; still this has little or no effect.  The detailed mechanism for diffusion could change, but if it remains Planckian, there would be scant evidence of a change of mechanism in the   resistivity.

\subsubsection{No sharp rise at low T}
The  $T^4$ rise in resistivity seen in normal metals below the Debye temperature will not apply, because the diffusion constant is already ``pinned'' at the Planckian rate as ever shorter modes awaken at higher temperature,  making scant difference. Indeed, the  $T^4$ rise is not seen in the strange metal regime, or if present in parallel is overwhelmed by the fast rise T compared to T$^2$ or T$^4$ at low T.

\subsubsection{Mott-Ioffe-Regel bypass}
The Mott-Ioffe-Regel resistivity rollover is bypassed in the strange metals for the following reason: the diffusion, responsible for the transport, is Planckian both before and after the MIR limit is passed, so the resistivity continues to rise   as $T$.

\begin{figure}[htbp]
     \centering
     \includegraphics[width=0.90\linewidth]{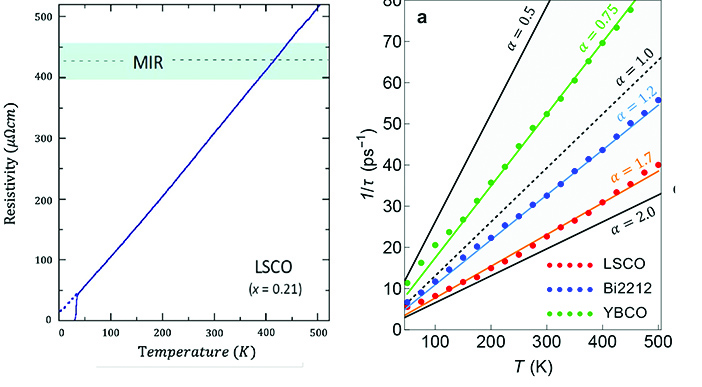}
     \caption{ Resistivity of three strange metals as a function of temperature, according to the wave-on-wave (WoW) approach. On the left, we note the  failure of the strange metals (here  of LSCO) to live by the Mott-Ioffe-Regel (MIR) in high-temperature resistivity rollover, which ``should'' have happened around the temperature of the blue band at the upper left. On the right, we see there is no also MIR rollover in our calculations, a fact we assign to WoW keeping electrons Planckian well past the distance and time of a mean free path~\cite{PNASpr}.  Also on  the  right, for parameters near optimal doping for LSCO. Bi2212, and YBCO, we see near-perfect linear resistivity at near Planckian rates down to about 10K~\cite{PNASpr}.  }
     \label{fig:alhunLinear}
 \end{figure}

From figure~\ref{fig:alhunLinear} of reference \cite{PNASpr} it is clear that linear resistivity at the correct Planckian slope prevails in the three strange metals investigated.  
\subsubsection{Planckian diffusion}

The diffusion constant $D$ in each of the three cases was close to $\hbar/m^*$.  Each of the deformation  potentials, if frozen, would have localized the electron, so again we have broken the Anderson localization only to find its ghost, Planckian diffusion.

   \subsection{ Electrons on a solid hydrogen surface} 
   \label{solidH}

\begin{figure}[htbp]
    \centering
\includegraphics[width=0.45\textwidth]{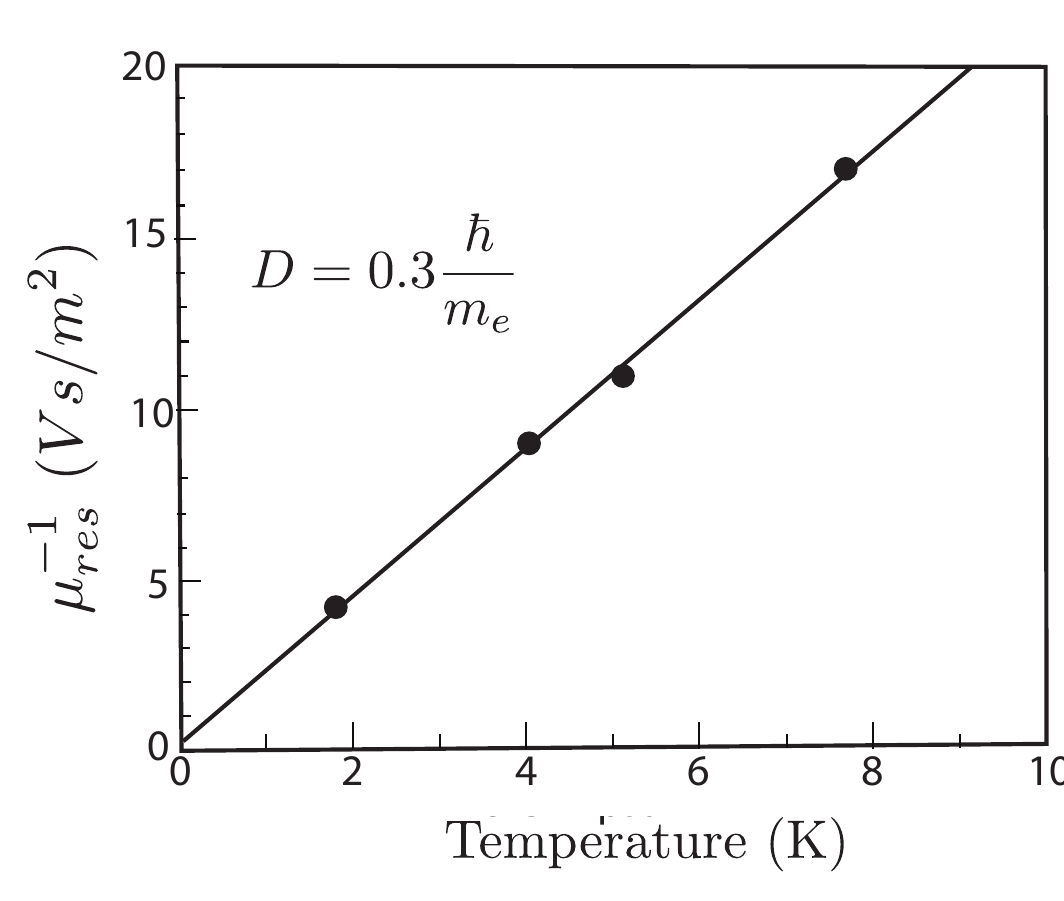}
    \caption{A re-drawn plot of the inverse residual mobility as a function of temperature, from Adams~\cite{adams_conductivity_1992}. The solid line is the best linear fit to the data. The slope was  interpreted by the author as  $D = 0.3\ \hbar/m_e$  }
    \label{hdata}
\end{figure}

We now refer to the experiment   by Adams, and Adams and Paalanen~\cite{adams_localization_1987,adams_conductivity_1992},  measuring  the mobility of nondegenerate electrons on a 2D solid hydrogen surface, perturbed by adsorbed Helium atoms.  The Helium atoms  release the electrons from bondage. Planckian diffusion $D\sim\hbar/m_e$ emerges, down to the lowest temperature measured, which is just below 2 K. The mobility is measured by magnetoresistance in a Corbino disk geometry (figure~\ref{hdata}).

Electrons are deposited on the solid hydrogen  surface, where they remain.  The  surface is rough and probably stepped, leading to localized quasi 2D states of vanishing mobility.  Helium is then co-deposited on the surface.   The  remaining gaseous fraction of Helium   is pumped out, and measurements  begin. It takes some hours for the surface Helium to escape. The Helium is certainly itinerant, and travels atop the hydrogen layer,  remaining fluid and mobile.  It supplies a time varying field to the electrons below, in an analog of a deformation potential,  disrupting the electron localization. If it is disrupted in a time shorter than the time to establish bound states, it does not matter if it is disrupted any faster, the diffusion will still be Planckian.  As long as some Helium remains on the surface, the electrons can diffuse and a nearly fixed Planckian mobility prevails.  This comports with the lack of  dependence on the strength of the disturbance in the diffusion constant expression, $D\sim\hbar/m^*$.

This experiment     emboldened us to believe the Planckian diffusion  we were seeing in WoW is  truly ubiquitous. 
It helps point the way to linear resistivity down to 0 Kelvin. 

Thus the criteria for nondegenerate universal Planckian diffusion  are met  under a temperature-independent diffusion constant  $D\sim \hbar/m_e$. The effective mass is very close to the bare electron mass.  The Planckian diffusion leads to linear resistivity with temperature, here in the 1-10 K region as seen in figure~\ref{hdata}.

Only a fraction of the medium (10 \% in our example) actually has to be on the move to break localization and transition to Planckian diffusion. This  lends support  to the idea that Planckian dissipation is universal when a localized particle or carrier is released by a very wide variety of  time dependencies of the localizing potential. It is a subject for future investigations to discover what happens as the collisions with a perturbing gas become inelastic and decohering.

\subsection{ Planckian diffusion from different moving impurities}
\label{impurGas}
With an eye to what might be happening near 0 Kelvin in the strange metals, we are beginning to question whether the strange metal landscape is nearly as flat and quiet as the thermal deformation potential alone would suggest, at very low T.
For example, there may be polarons, or charge density fluctuations, or simply the ``wakes'' of other electrons, caused by their passage through the area, like so many motorboats on a lake that would have been still without them.  There is also the Coulomb repulsion of  nearby passing electrons at the Fermi energy.  One or more of these effects can be causing Planckian diffusion near 0 Kelvin, even in the absence of thermal agitation built into the deformation potential.  Planckian diffusion is universal and easy to reach, and with it, goes universal T-linear resistivity. If the mechanistrange metal leading to Planckian diffusion switches with temperature, there would be scant trace in the resistivity.

We model the potential field felt by a charge carrier as a collection of moving potential bumps, representing a polaron gas\cite{ghost}. Each bump is initialized with a random position and velocity, based on the Maxwell distribution. A Gaussian wave packet is launched into this moving potential field. We use the split-operator WoW algorithm to solve the time-dependent Schrödinger equation. By calculating the Mean Square Displacement (MSD) at each time step, we derive the diffusion coefficient using the following formula:
\[D =  \lim_{t\to\infty} \frac{1}{2d} \frac{\partial MSD(t)}{\partial t}\] 
where $d=2$ is the dimensionality of the system. The long-time limit ensures that initial conditions are forgotten.

\begin{figure}[htbp]
    \centering
    \includegraphics[width=1\linewidth]{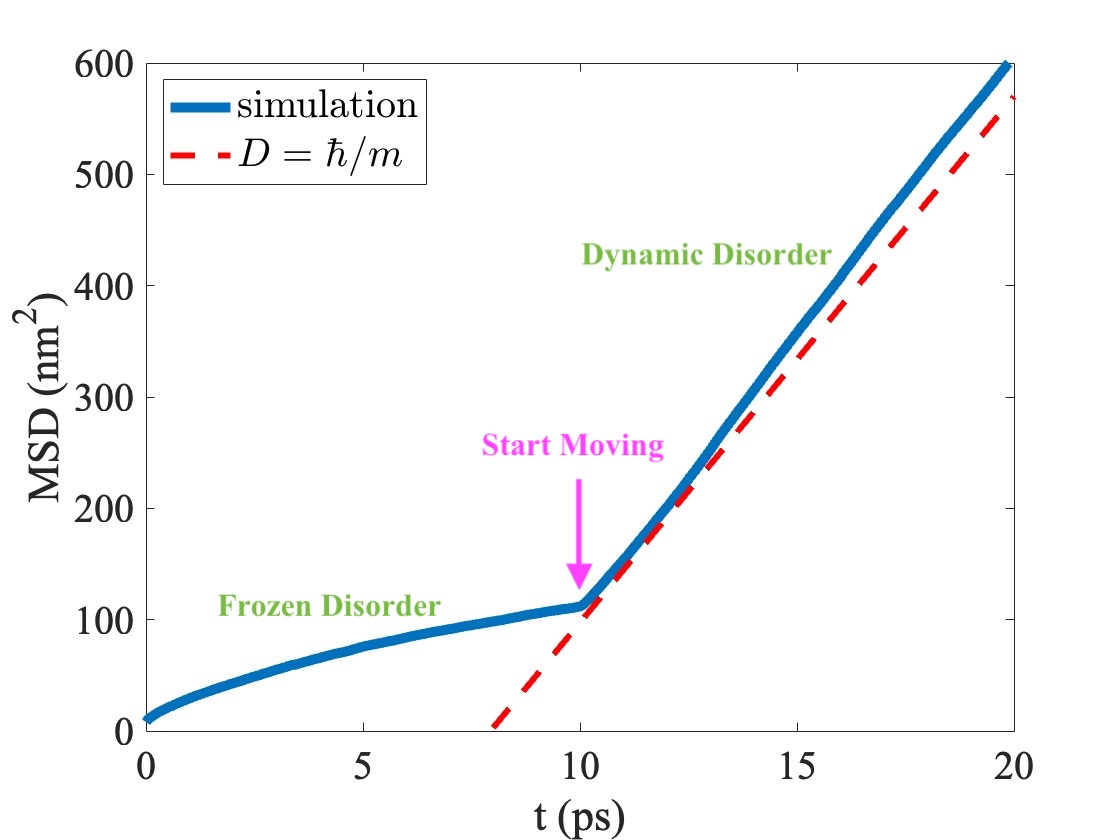}
    \caption{Electron transport in a field of randomly distributed frozen   impurities, which begin random motion after 10 ps.   After launching a Gaussian wavepacket,  diffusion is slowing down due to Anderson localization from 0 to 10 ps. After 10 ps, the impurities are set in random motion, leading immediately to Planckian diffusion, verified by the linear behavior of MSD against time, which has a slope near $ \hbar/ m$ diffusion\cite{ghost}.}
    \label{frozenpolar}
\end{figure}

Initially,   the polaron-like impurities are  frozen in position,   Anderson localization of the wavefunction is setting in for the first 10 ps, as expected for a system with static disorder. After 10 ps,  we assign finite velocities to the impurities, Anderson localization is destroyed, giving way to diffusive with a diffusion constant close to $\hbar/m$.2(figure~\ref{frozenpolar}).

We also found that if even a  fraction of moving defects, with the remainder still fixed,  can break the Anderson, leading to Planckian diffusion. Remarkably, as little as 10$\%$ of the defects in motion is enough to induce this transition.

 \subsection{   Dynamical  Planckian diffusion model} 
\label{Planckian}

\begin{figure}[htbp]
\centering
\includegraphics[width=0.35\textwidth]{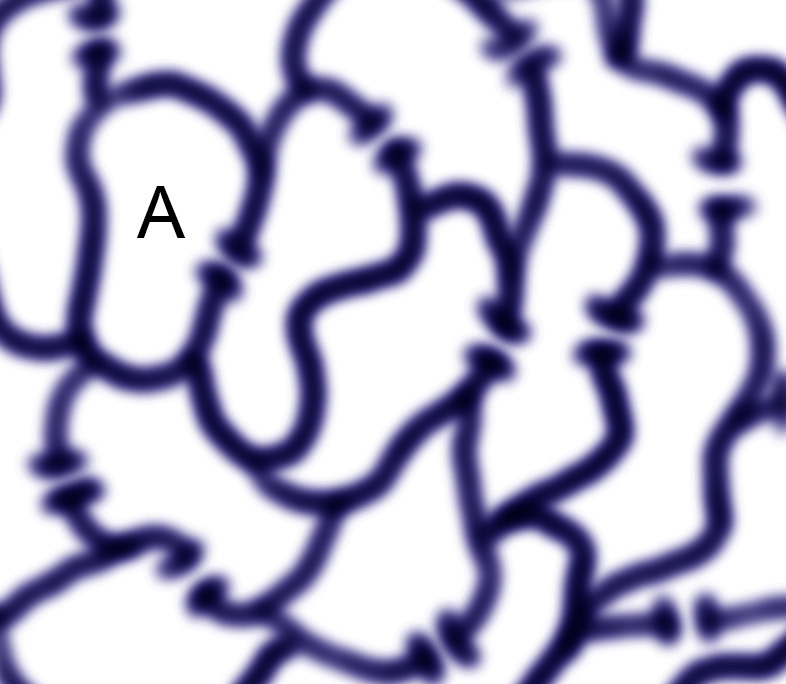}
    \caption{ Chambers of area roughly $A$ are connected by single-mode quantum channels or contacts; these may slowly open or close between chambers. This gives rise to Planckian, quantum diffusion, with $D\sim \hbar/m^*$, independent of $A$. There is no defined temperature, and indeed the Planckian idea should not depend on there being anything thermal, although it is compatible with thermal systems, where Planckian diffusion  implies the Planckian speed limit, $\tau = \hbar/k_B T$, through an Einstein relation. Unfortunately this rule does not supply its own  insight as to what process $\tau$ is supposed to represent.} 
    \label{Planckian2}
    \end{figure}

We present a simple, dynamic model  that   obeys Planckian diffusion $D \sim \hbar/m^*,$    leading  (if any temperature is supposed, which is not necessary) to perfectly  linear in temperature resistivity down to  0 Kelvin. The linearity in temperature comes from concentration gradients induced by applied fields, together with temperature-independent Planckian diffusion, not from any microscopic or mechanistic sensitivity of the dynamics to temperature. With the gradients established, Fickian diffusion reigns, causing transport, and linearity prevails. 

The model is a cousin of the Thouless model\cite{Edwards_1972} of the transition from diffusive to localized transport, but it differs in important ways: the way of reaching the boundaries is quasi-ballistic, not diffusive, and most of the boundary remains opaque.   The model was inspired by observing quantum acoustic wave-on-wave (WoW) simulations, where we see electrons roam around at the Fermi velocity in rough,   constantly morphing zones (preventing Anderson localization) and sometimes escaping to an adjacent zone. As they escape, the existing basin starts to drain as probability flows to an adjacent one, often  without visible flow in between.  We call this ``ghost walking''.

We assume a collection of adjacent leaky boxes which are ballistic chaotic billiards.  The limiting factor to get out is probability of being at the boundary, which is proportional to $1/A$, where $A$ is the box area, together with the flux out after arriving  at the boundary, which depends on arriving at an open channel. 

In the many electron picture,  chambers of roughly of area A are filled up to something like a common Fermi level, with variations in the top-filled level from box to box. We suppose there are many more states near the Fermi level in each box than carriers, resulting in  effectively nondegenerate electrons or carriers.


The boxes become smaller with increasing temperature and/or  electron-lattice coupling, but the diffusion constant will turn out to be independent of the box area. 
The electrons escape from one box to the next, randomly in any direction.   The escape rate from the channels, given  below, is   distinctly quantum.  

 We  seek the diffusion constant $D$ of this model.    For a 2D box of area $A$, the density of states is 
$$\rho = 2 \pi m A/h^2.$$
A single channel ``leak'' allowing escape from a chaotic box gives a decay lifetime that broadens the levels to the single  level spacing,  
$$\tau=\hbar \rho,$$
or $\tau = m A/h.$ So, if escape from one chamber to the next leads to a 2D random walk, with stepsize $d=\sqrt A$, and time between steps $\tau = m A/h,$ the 2D random walk diffusion constant $$D = d^2/4\tau =  2\pi \hbar/4m \sim \hbar/m,$$ independent of temperature, deformation potential constant, or confinement area $A.$ 

There is a kind of   quantum criticality implied when the level spacing equals the level widths
\subsection{ Drude peak displacement}
	The Drude peak in ordinary metals is a sharp rise in light absorption of the material at low frequency, resulting from nearly free carriers in the metal. At low frequencies  the electrons are slowly dragged back and forth, as in a   DC field; the absorption is basically due to DC resistivity.     The Drude peak is absent in insulators or semiconductors without a high density of free electrons. 

\begin{figure}[htbp]
\centering
\includegraphics[width=0.35\textwidth]{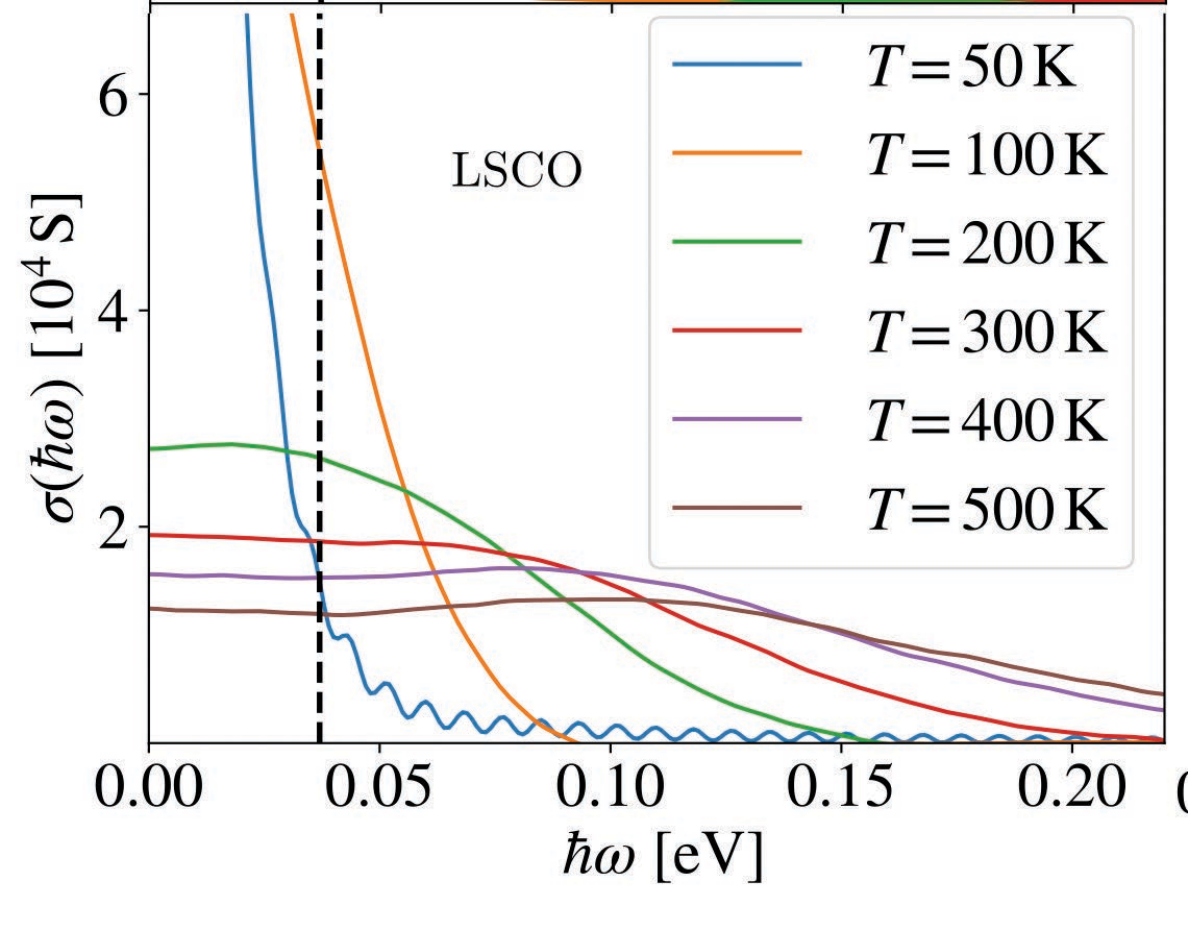}
\caption{ Computed optical conductivity for LSCO at different temperatures, resolved within the static and dynamical potential landscapes averaged over 100 and 10 realizations, respectively. The back dash line marks the Debye frequency of the  material below which the frozen potential assumption breaks down. A key dynamical effect is the saturation of conductivity in the regime $\omega \lesssim \omega_{D}$. However, even if the potential is frozen, we find, the optical conductivity peak shifts from the Drude-peak    situated at $\omega = 0$ to higher energies,   broadening as the temperature increases.}
\label{Fig:frozen_vs_dynamical}
\end{figure}

    Fratini and  Ciuchi considered Drude peak displacement in this and similar contexts and assigned it to unspecified slow bosonic fluctuations\cite{infrared}.  We specifically modeled  the exiled (see taboo above) lattice vibrations and found that they were responsible, after all:
    
In a recent highlighted PRL~\cite{prlDDP}, we addressed this question using WoW techniques. See figure~\ref{Fig:frozen_vs_dynamical}.
The quantum-acoustical representation reveals a properly displaced Drude peak hiding in plain sight at higher temperatures within the venerable Fröhlich model: the optical conductivity obtained from thw WoW simulations the Kubo formalistrange metal exhibits a finite frequency maximum in the far-infrared range, while  the near-d.c. conductivity is suppressed. 

An electron sees a Gaussian random time dependent deformation potential landscape;   at 50 K the  thermal deformation potential itself is strong enough to cause Planckian diffusion of the electron, with $D=\hbar/m^*$, where $m^*$ is the effective electron mass.  

The interpretation is that at the deformation potential becomes formidable, the Fermi level electrons are no longer dragged with friction across the landscape, reducing absorpion at very low frequency.  At slightly higher frequency, the electrons can absorb photons, promoted to low energy excited states lying just above the Fermi enerfy.

 Below 50 K, we suspect the deformation potential may get  ``assists'' from other disturbances to the  potential seen by the electrons, such as polarons or CDW. 

At higher T, if the mobility becomes diffusive, electrons are much less prone to being affected by the field, reducing field induced dissipation (see figure \ref{nomobility}). However something like Franck-Condon transitions become possible, at higher frequency.  Thus the 0 frequency peak diminishes and the absorption migrates to higher frequency, at higher temperature: Drude peak displacement.

Our WoW calculations were initiated with bound states of a frozen deformation potential, The deformation potential was then allowed to move normally and the conductivity was calculated numerically.

We focused on three prototypical compounds classified as strange/bad metals, namely LSCO (shown here), Bi2212 and Sr\textsubscript{3}Ru\textsubscript{2}O\textsubscript{7}.  By computing the velocity autocorrelation for all time steps, we numerically determine the optical conductivity in the case of a dynamical lattice disorder field within the Kubo formulation. Figure\ref{Fig:frozen_vs_dynamical} show that the displaced Drude peak emerged at higher at higher T as desired. The movement with temperature of the broad new peak at higher frequency evolved as it should.  More details may be found in reference~\cite{prlDDP}.

.
 
\subsection{Polarons, CDW, and the linear approach to 0 Kelvin}

\subsubsection*{ Polaron formation}

Of course, we cannot carry out a full many-body calculation
with thousands  of lattice modes into the picosecond time scale.  Mean field methods are a natural resort. 

There are a few things known to be suspect within mean-field. One of course is an improper accounting of entanglement. Amplitudes for events that take place in different places at different times are not interfering as they should.  However, it is easy to speculate that this does not matter much, since such interference is probably moot for bulk questions like how far has an electron moved, given all the opportunity for self averaging.

We have been able to check the efficacy of mean field in our case, through  the Feynmann influence functional approach; see section \ref{path}.  The strange metals are likely in the strong coupling regime where mean field is known to hold~\cite{FeynmanInfluence}

We were initially surprised to see  polarons form spontaneously in our mean-field, WoW simulations. Such spontaneous polaron formation was shown in figure~\ref{polaron3psec}. We again call attention to the fact that this entire scene is the result of a single many body coherent state. One might have expected to create a polaron by hand and see if it is stable within mean-field, at best. Instead, acoustic polarons form spontaneously, of approximately the right size and energy.  They form suddenly, sending out a Tsunami of wave energy at the sound speed. All the while, the total energy, electron plus lattice,  is very nearly conserved\cite{aydin_polaron_2025}.

More serious is the question of the local electron probability amplitude, since the effect of the electron locally on the lattice goes as $|\psi|^2$.   However a recent PNAS paper ~\cite{aydin_polaron_2025}, allays these fears, showing how robust the numerical results are.  Even the polaron formation dynamics is convincing: after a picosecond or so at temperatures under about 20 K, they form suddenly, although we do not yet understand just what happened the moment before.
We can   be fairly confident that a similar polaron formation scenario would apply in a full many-body implementation.   


\subsubsection*{ CDW formation}

It is expecting too much to think that one mean field electron roaming a 40 nm$^2$ window should cause CDW or polarons at all, much less at the right temperature and forming with the right energy and geometry. Yet we see very plausible CDW and polarons form as we watch the calculation unfold.  

Some good fortune helps to explain this. First, because of periodic boundary conditions, we have an effective carrier density of one electron per 40 nm$^2$ in a 40 nm periodic window. As amplitude leaves the right side of the window, it re-enters the left encountering a landscape it created earlier, but it may as well have be caused by passage of another electron.
\begin{figure}[htbp]
    \centering
\includegraphics[width=0.45\textwidth]{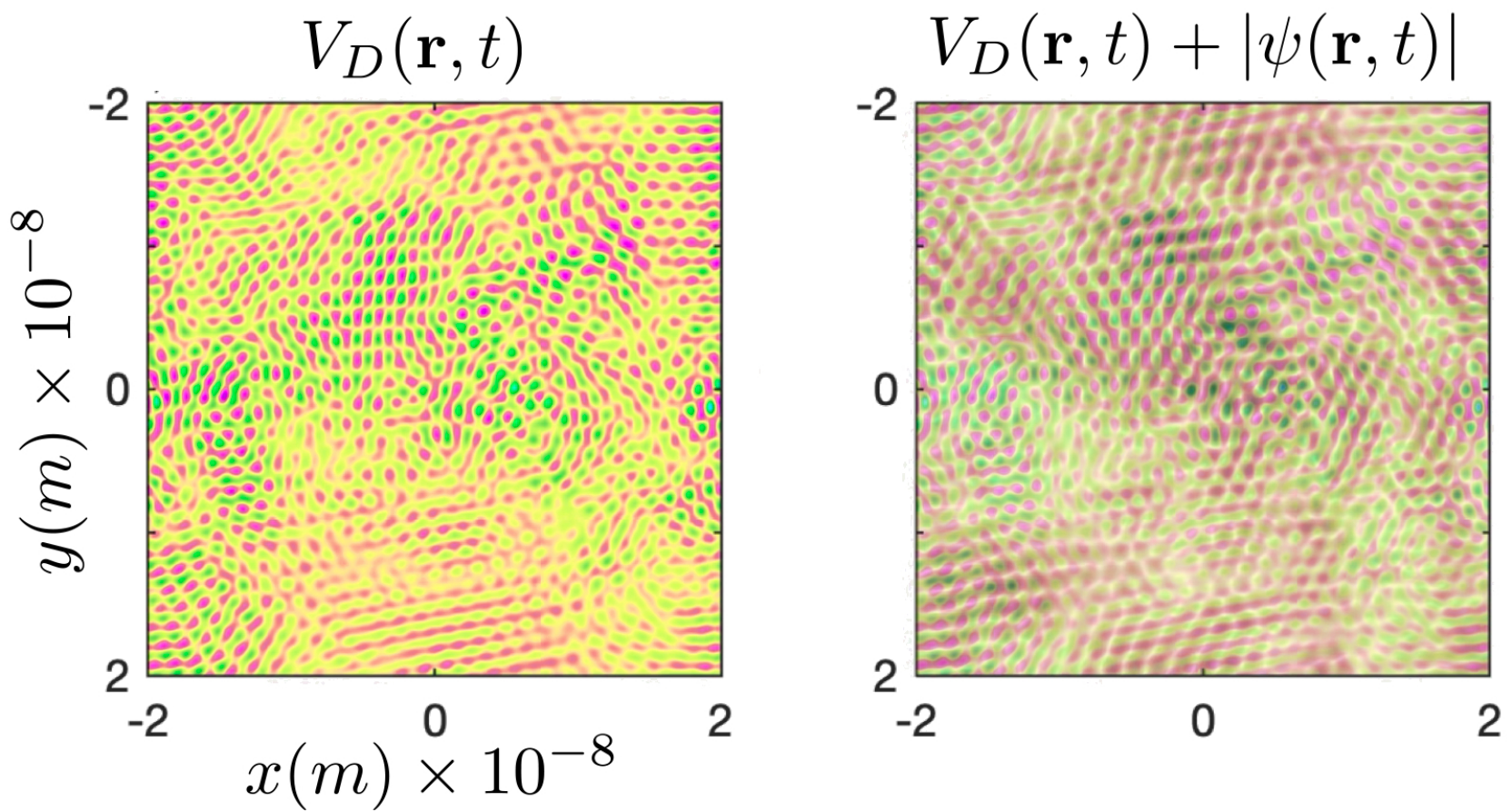}
\caption{Spontaneous CDW formation in a WoW simulation after about 6 psec. On the left is seen the low temperature (3 K) deformation potential,  the density wave in color showing positive  potential, red, and negative, blue.  On the right, the deformation potential is again shown in the same colors, with $\vert \psi({\bf r},t)\vert$ overlain in grayscale revealing a charge density wave.  Inspection reveals a nearly complete alignment of the two. The patterns move  and morph slowly, together. }
\end{figure}

Because polarons or CDW and nascent fluctuations are not phonons, we replace the idea of ``electron-phonon'' interactions,  with ``vibration-electron,''  or better, ``vibronic'' interactions.   The concept of vibronic interaction has been in use for decades in molecular physics and implies an electronic response to vibrational motion, and {\it vise-versa}, be it adiabatic or nonadiabatic. It maay be that the elusive understanding of the cuprates has not been  in finding  an emergent e-e fluid, but rather in developing a proper account of vibronic interactions, including polarons, CDW,  and their associated fluctuations.

Typical $E_d$ in equation \ref{Ed} is  10-30 eV.  This is quite large and may put us in good stead for the mean field approximations, which are proven to be exact regarding some important properties in the strong coupling limit~\cite{leopold2021derivation,leopold2021derivation,leopold2022landau}.   

In filling up the Fermi sea level by level, the  total energy shift is the sum of individual level shifts up to the Fermi energy. This is worth noting because in a real sense, the electron-lattice coupling is strong at any temperature for the strange metals. For example,  polarons, which are an extreme outcome of electron-lattice coupling,   typically form {\it below }50 K  in our wave-on-wave simulations described below.  They reflect strong vibronic coupling present even at 0K. The intrinsic electron-lattice coupling does not diminish with lower temperature. These are obvious enough comments, but if electron-phonon coupling is strong enough, and the Fermi energy and velocity are low enough, strange things can happen near 0 Kelvin. It is risky to assume the landscape that an electron sees near 0 K in the strange metals is serene, and incapable of scattering,  just because the thermal part of the deformation potential has gone nearly silent. The engine of Planckian diffusion leading to linear with temperature resistivity may be alive and well. 

\subsubsection* { Primacy of Planckian diffusion}
  Planckian diffusion is the engine behind  the Planckian timescale $\tau\sim\hbar/k_B T$. The latter  is less general because it  applies  only  to thermal systems, and      {\it follows} from Planckian diffusion.  This casts a new light on discussions   regarding what process the  Planckian timescale  represents, because the foundational behavior lies elsewhere:  Planckian diffusion.  No necessity of defining a process with a timescale. 
  Such concerns can add an unnecessary layer of uncertainty if the models used to define the timescale are unavoidably imprecise.  An example is the collision time $\tau$ in the Drude picture.  In the strange metal region at least, there   are no clearly delineated collision events, rather a continuous strong interaction. 
  
  But Planckian diffusion is  ubiquitous, and perfectly well defined, with  a constant $D$, in two dimensions, 
  \begin{equation}
D = \lim_{t \to \infty} \frac{1}{4} \frac{d}{dt} \langle r^2(t)\rangle =\int_0^\infty \langle v(0) v(t) \rangle dt.
\end{equation}
Together with the expression for the mobility $\mu$, again a ``safe'', model independent quantity,
\begin{equation}
\mu = \frac{q}{k_B T} \int_0^\infty \langle v(0) v(t) \rangle dt
\end{equation}
we have 
\begin{equation}
D = \mu k_B T,
\end{equation}
still free of any models for transport. 
  Now, a temperature has crept in, but not a timescale. The temperature can be attributed to Einstein and the fluctuation-dissipation theorem.  However,   detailed balance may not hold for diffusive transport, where chemical potential gradients intrinsically cause entropy production.

  In the standard Einstein-Drude  picture, if it applies, 
  $$D = \frac{\tau}{m^*} \ k_B T,\label{linear3}$$
then in one line, with Planckian $D = \hbar/m^*,$ we get the speed limit $\tau = \hbar/k_B T.$ Planckian diffusion is the foundation, the rest depends on models for $\tau,$ if one casre to make them. 

The diffusion $D\sim \hbar/m^*$, on the other hand, offers a direct experimental handle, i.e. the spatial spreading of probability, without requiring assumptions about what caused the diffusion.

 An insight into this ubiquity emerges  in the model given section \ref{Planckian}.   There is no mention  there, nor is there necessarily a role, for a temperature or Planckian timescale $\tau \sim \hbar/k_B T$. 
 
\subsection{Discussion and implications}

We end with speculations based on what we have learned so far.

 \subsubsection*{ Linear  resistivity  to 0 K}
\label{linear}
One of the key features making the strange metals strange is perfectly linear in temperature resistivity over orders of magnitude in temperature, down even to perhaps 0.1 Kelvin, if superconductivity has been suppressed by a magnetic field. This holds in the strange metal phase of figure \ref{phasedia}, and the linearity is seen in figure \ref{fig:alhunLinear}, though not down to 0.1 K. More than that, the slope of the resistivity vs. temperature is limited by a Planckian bound,  in the form of a Planckian speed limit.  


  Since we are in a speculative mode, we discuss a possible electron-lattice soup forming in the strange metal regime, even at low T, and dressing up electrons (or really holes) to act nondegenerate.

\begin{figure}[htbp] 
   \centering
\includegraphics [width=2in] {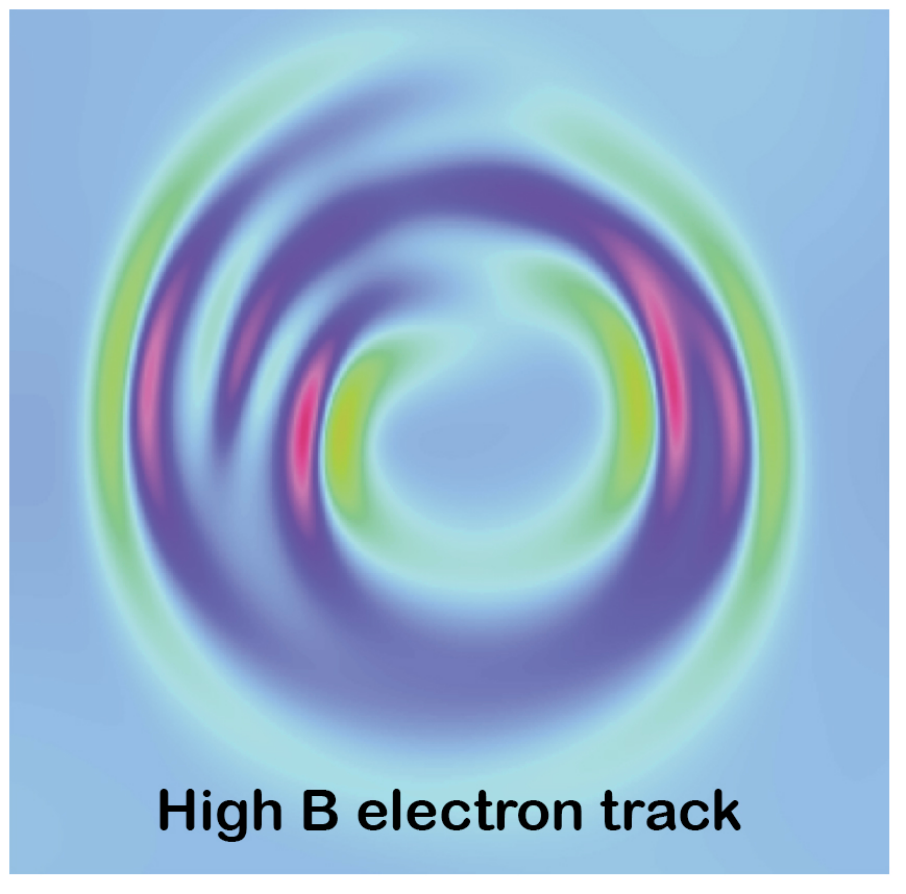} 
   \caption{In a wave-on-wave simulation,  with normal strange metal parameters, a pliable, interactive electron-lattice interaction in the low Fermi energy optimal doping strange metal regime of the cuprates is seen. A modest deformation potential constants of 10 eV was used. Only the deformation potential is shown, and the landscape is the result of a single, multidimensional coherent state of the lattice. Here, an electron circling at a high magnetic field perpendicular to the plane has made just over one Landau orbit, leaving behind a ``speedboat'' wake in the low $T$, initially quiet deformation potential. If many electrons were doing this in equilibrium, the potential would be roiled,  even near 0 K, giving back as much energy as it receives in a dynamic equilibrium. This regime is strongly suppressed in an ordinary metal with high Fermi energy and velocity. The point is, the landscape is anything but flat. In the energy domain, this corresponds to the debate over rippling of the lattice at 0 K.}
 \label{fig:soup}
\end{figure}

Electrons at the Fermi level in the cuprates carve  backaction grooves in the deformation potential, much like a speedboat creates waves (see figure~\ref{fig:soup}). We assume holes do this too.    Supposing many speedboats on a lake with no wind, i.e.  no thermal  agitation, the waters are nonetheless anything but calm. The deformation potential is  fluctuating, and acting back on the electrons, in equilibrium.  Although the electron-lattice back action can cause CDW in other regimes, we do not need these, nor polarons,  to explain the rough landscape that even cold temperatures present to the electrons. With such roiling, Planckian diffusion becomes plausible even at very low T. We suggest this as one possible explanation for even very low T linearity.                                         This scenario  has to occur in equilibrium, which means there is a kind of active electron-vibration  soup formed, with electrons and holes  in constant, two-way exchange of energy with the lattice. Electron energy would not be separable from the lattice in this strongly correlated soup. What is  the compressibility  of such a soup, as a function of T?

 \subsubsection*  { \it Becoming strongly correlated}

We suppose that electrons are indeed strongly correlated in the strange metals, as is so often stated, but there is a big caveat: it is an effect of strong  vibronic interaction~\cite{strangemetal1}, This happens through electron back-action on the lattice, {\it regardless of the temperature}.  Since other electrons interact with the same vibronic soup, there are lattice mediated electron-electron interactions.

An example is given by two nearby polarons, each of course confining an electron. In mean field, both polaron pockets are described by a single coherent state in many coordinates; the lattice is mediating and strongly correlating two electrons. If screened Coulomb repulsion is accounted for, and perhaps mutual attraction of the lattice pockets, a many body correlated state is established.  We suppose that related things  are going on without polarons present,  in the looser sense of a soup, 

 

\subsubsection* { Lorenz ratio}
 
 Instead of electrons leading the way to thermal transport in the usual pure metal Fermi liquid thermal conductivity,  in the strange metals it seems   that the thermal deformation potential and back-action lead the way, in a kind of extreme phonon drag, with potentially important consequences for the Lorenz ratio~\cite{strangemetal1, planckianmetals}. Indeed the scaled Lorenz ratio is above 1 for strange metal-like behavior, and below 1 for ordinary metals \cite{strangemetal1}, indicative of supremacy of the thermal conductivity in the strange metal phase.  The Lorenz ratio is on our list of future targets of opportunity.


We find large ranges of parameter temperature (if any temperature applies), coupling to the medium, speed of the medium,  nature of the medium randomness, and even what percentage of the medium is in motion, all with $D =\alpha \hbar/m^*$ and $1/2<\alpha <2$.
 The independence of $D$ from the temperature and electron-lattice coupling constant is remarkable.   It can be used to derive linear in T resistivity in the case of diffusion transport (vanishing Drude-Einstein drift mobility), assuming nondegenerate carriers.

\subsubsection*{ A new ubiquity}

The Planckian $D$ is not so much of a bound as it is ubiquitous; it {\it can} be violated in either direction, e.g. under extremely slow or extremely fast evolution of the medium. It seems clear that if the potential ``scene'' is locally and significantly altered before the time required to set up the delicate return-path coherence needed for localization, then localization will be compromised. That surely must lead to  diffusion. Our advance is that it is Planckian diffusion - and a new ``universality," with the quotes allowing for exceptions.

Planckian diffusion is appropriately called a ghost,  having some of the qualities it had before death, but now it is itinerant and ephemeral. The agent disrupting the medium can be many different things, adding a new dimension of ubiquity.


In the case of   electrons on solid hydrogen, the disrupting agent is the presence of mobile helium atoms. Planckian diffusion in a moving random medium is the direct product, so to speak, of two universalities: the means of localization and the means of release.

\subsubsection*{ Isotropy and absence of hot spots}
An important ramification of   quantum diffusion is  its isotropy, which was emphasized by Grissonnanche {\it et. al.} in 2021~\cite{Grissonnanche2021}, and taken as evidence against ``hot spot'' scenarios.  Quoting, ``we find … a T‐linear scattering rate that saturates the Planckian limit … Remarkably, we find that this Planckian scattering rate is isotropic, i.e. it is independent of direction, in contrast with expectations from ‘hot‐spot’ models." 

  The reason for the isotropy is the    universality we are now discussing: the diffusion, having neared  a Planckian limit, is uncaring about details of the diffusing potential, and yields the same diffusion in any direction. Or put another way, the hot spots may be present but they fail to affect the diffusion very much.  This ``uncaring'' property has colored  many of the arguments against ``phonons'', i.e. the lattice, when   effects like isotope substitution did not produce  the expected effect.  We know, for example, that the deformation potential strength can vary many fold without much effect on the diffusion, which remains $D\sim \hbar/m$
In the WoW simulations, however, coherence at least in the form of  Planckian resistivity, which is purely a quantum phenomenon,  outlives the mean free path. In fact the very concept of a mean free path is moot under the strong field diffusion that controls the strange metal regime. There is no remnant of Drude or any ballistic motion. In terms of a modern literature, the transport is ``incoherent''  meaning transport cannot be explained by quasiparticles; and instead is described by diffusion constants, and Planckian scattering.  Scattering and resistivity continue on their linear rise with temperature. With parameters chosen to reflect three strange metals at optimal doping, no deviation was seen  at high temperature, just as in the experiments~\cite{PNASpr}.

\section{Summary}
The quantum acoustic perspective given here involves a time dependent, quantum formalism not found in solid state theory texts. However we are far from being out on a theoretical limb: the choice of the coherent state approach to the lattice leads us  to this place  is  extremely well established in an adjacent field: quantum optics.  

 Arriving at the  wave-on-wave (WoW) formal and numerical approach is  almost not a choice; it is  virtually forced upon us in the coherent state representation, although we are willing advocates of this beautiful and intuitive representation of quantum mechanics.

All of the classic normal metal behavior has been affirmed within quantum acoustics.  There are a number of strange metal phenomena that have already yielded to the quantum acoustic approach:   Mott-Ioffee-Regel bypass, Drude peak displacement, spontaneous Fr\"olich polaron and CDW formation, Planckian speed limit resistivity linear in T.  We have in our sights, without making an airtight case as yet, very cold linear in $T$ resistivity, pseudogap physics, and the backward sloping pseudogap strange metal boundary.  We are pursuing plausible scenarios in each case, as sketched above.


 Planckian diffusion is ubiquitous, and the diffusion constant $D=\hbar/m^*$  over huge parameter ranges remains no matter what agent is behind the stirring of the medium (but it must be stirred). Faster or more vigorous stirring has little effect on the Planckian rate.  So strange metals  can change temperature, or doping,  yet have almost exactly the same diffusion, and the same linear resistivity in T.  

 We again remark that this work started with the venerable workhorse   Fr\"olich Hamiltonian, and never had to deviate from those beginnings. We have shown that much strange metal physics can found in contained with that model, and probably more will be found.  The Hamiltonian itself stops at the lowest order electron-lattice strain expansion, yet when that interaction is allowed to act continuously,   coherently, and to infinite order, the treasure trove of physics it contains is revealed. 


\section{Acknowledgements}
This work was supported by the U.S. Department of Energy under Grant No. DE-SC0025489. In the early stage of this work, we benefitted from the support of the former NSF Center for Integrated Quantum Materials (CIQM).  We thank  Harvard University's Dean's Fund, and the Harvard Quantum Initiative for support. We thank Prof. Supryo Datta for his multiple insights and consultations. We have benefited from scientific discussions with Prof. Brad Ramshaw and Prof. Carlos Trallero.

            
            

\section{Appendix: Coherent state formulation}

We provide some support for the introduction of a powerful  path integral approach to quantum acoustics in section~\ref{path}.
We begin by considering a composition system of $\mathcal{H}=\mathcal{H}_S+\mathcal{H}_B+\mathcal{H}_I$. It entails a generic quantum system of charge carriers $\mathcal{H}_S$ that interacts with a lattice $\mathcal{H}_B$ modeled as a bath of harmonic oscillators. By denoting the normal coordinates of a normal mode as $\boldsymbol{X}_{\boldsymbol{q}} =\left[\begin{matrix} x_{\boldsymbol{q}} & p_{\boldsymbol{q}}  \end{matrix}\right]$, the lattice bath is defined as 
\begin{equation}
    \mathcal{H}_B = \frac{1}{2}\sum_{\boldsymbol{q}} \omega_{\boldsymbol{q}}\Big (\boldsymbol{X}_{\boldsymbol{q}}\cdot\boldsymbol{X}_{\boldsymbol{q}}^T \Big).
\end{equation}
and it is linearly coupled to the system $\mathcal{H}_s$ in the following way
\begin{equation}
    \mathcal{H}_I = \frac{1}{2}\sum_{\boldsymbol{q}} \Big( \boldsymbol{g}_{\boldsymbol{q}}(\boldsymbol{r}) \cdot \boldsymbol{X}_{\boldsymbol{q}}^T \Big)
\end{equation}
where $\bm g_{\bm q}(\bm r) = \left[\begin{matrix} g^0_{\bm q}(\bm r) & g^1_{\bm q}(\bm r) \end{matrix}\right]$ consists of model-specific functions depending only on the system coordinates $\boldsymbol{r}(t)$. 

In a fashion similar to quantum optics~\cite{walls2007quantum, scully1997quantum}, we derive the corresponding quantum master equation that governs the exact dynamics of the reduced density matrix $\rho_S$ of a system of interest, starting from the path-integral formulation. Here, we only outline the key definitions and steps, while the more complete derivation can be found in the supplementary material.

We start by assuming the initial density matrix $\rho$ of the composed system is separable, i.e. $\hat \rho(t=0)=\hat\rho_S\otimes \hat\rho_B$. Such a state is not easily found in nature, but is a common starting point for theoretical analysis. Generalizing the formalistrange metal to other initial states by adding an additional path integral over imaginary time is an avenue for further research\cite{weiss2012quantum}. Employing the independence of normal modes, entire lattice vibrations can be described as the product state of the coherent states of the normal modes, i.e., as a multimode coherent state, hence resulting in the following density matrix
\begin{equation}
\label{eq: coherent bath}
\begin{split}
    \rho_B( \boldsymbol{X},\boldsymbol{X}';\boldsymbol{X}^0)\\ = \mathcal{N} \exp \Bigg\{&-\frac{1}{2} \sum_{\boldsymbol{q}} \Big[ (\boldsymbol{X}_{\boldsymbol{q}}  - \boldsymbol{X_{\boldsymbol{q}} }^0)  \cdot \bar{\boldsymbol{A}} \cdot ( \boldsymbol{X}_{\boldsymbol{q}}  - \boldsymbol{X}_{\boldsymbol{q}} ^0)^T\\ &+ (\boldsymbol{X}_{\boldsymbol{q}}' - \boldsymbol{X}_{\boldsymbol{q}}^0)  \cdot \bar{\boldsymbol{A}} \cdot ( \boldsymbol{X}_{\boldsymbol{q}}' - \boldsymbol{X}_{\boldsymbol{q}}^0 )^T\\ &-2\boldsymbol{X}_{\boldsymbol{q}}^0\cdot \bar{\boldsymbol{Z}} \cdot ( \boldsymbol{X}_{\boldsymbol{q}} - \boldsymbol{X}_{\boldsymbol{q}}' )^T \Big] \Bigg \},
\end{split}
\end{equation}
where $\mathcal{N} = \pi^{-N/2}$ is a normalization constant and we have defined
\begin{equation}
    \bar{\boldsymbol{A}}
    = \begin{bmatrix}
        1 & 0\\
        0 & 0
    \end{bmatrix}
    \quad \textrm{and} \quad
    \bar{\boldsymbol{Z}}
    = \begin{bmatrix}
        0 & 0\\
        i & 0
    \end{bmatrix}
\end{equation}

Instead of taking this single multimode coherent state as the initial lattice bath, a thermal ensemble can also determined in accordance with quantum optics as
\begin{equation}\label{eq: thermal prob distr}
\hat \rho_T=\int P( \boldsymbol{X}^0)\hat\rho_B( \boldsymbol{X}^0)d \boldsymbol{X}^0
\end{equation}
with the probability distribution 
\begin{equation*}
    P(\boldsymbol{X})=\prod_{\bm q} \frac{1}{\pi \bar{n}_{\bm q}}\exp\left[ -\frac{\boldsymbol{X}_{\boldsymbol{q}}\cdot\boldsymbol{X}_{\boldsymbol{q}}}{2\bar{n}_{\bm q}} \right]
\end{equation*}
where $\bar{n}_{\bm q}$ is the average number of quanta in a mode according to the Bose-Einstein distribution
\begin{equation*}
    \bar{n}_{\bm q}=\frac{1}{\exp(\omega_{\bm q}/ k_B T)-1}.
\end{equation*}
Henceforth, we however focus on the multimode coherent state, but we want to emphasize that the following procedure nevertheless applies to the thermal bath as well. 

Given the initial state $\rho_S(t=0)$, the state of the system $\rho_S(t)$ at later $t$ is determined~\cite{weiss2012quantum} by the propagator $J$ as 
\begin{equation}
\rho_S(\bm{r}_f,\bm{r}'_f;t)=\int d\bm{r}_id\bm{r}'_i J(\bm{r}_f,\bm{r}_f',t;\bm{r}_i,\bm{r}_i',0)\rho_S(\bm{r}_i,\bm{r}'_i;0),
\end{equation}
where the subscripts $i,f$ stand for initial and final coordinates, respectively. The propagator is on the other hand given by the following path-integral.
\begin{equation}
J(\bm{r}_f,\bm{r}_f',t;\bm{r}_i,\bm{r}_i',0)=\int\mathcal{D}\boldsymbol{r}\mathcal{D}\boldsymbol{r}'\exp(i S_S[\bm r]-iS_S[\bm r'])\mathcal{F}[\bm r,\bm r'],
\end{equation}
where the effect of the bath is incorporated in the influence functional $\mathcal{F}$. 

In the the center of mass and fluctuation coordinates
\begin{equation*}
    \bm u_{\bm q}(t)=\frac{1}{2}\left[\bm g_{\bm q}(\bm r)+\bm g_{\bm q}(\bm r')\right]
    \quad \textrm{and} \quad
    \bm v_{\bm q}(t)=\bm g_{\bm q}(\bm r)-\bm g_{\bm q}(\bm r'),
\end{equation*}
the influence functional can be written as
\begin{equation*}
    \mathcal{F}[\bm r(t),\bm r'(t)]=\exp{\left[-iS_{\text{mf}}[\bm r'(t)]+iS_{\text{mf}}[\bm r(t)]-\Psi[\bm r(t),\bm r'(t)]\right]},
\end{equation*}
where we have identified the mean-field action $S_{\text{mf}}$ and the influence phase $\Psi$. By introducing an additional notation
\begin{equation*}
    \bar{\boldsymbol{L}}_{\bm q}(t'-s)=
    \begin{bmatrix}
    \cos[\omega_{\bm q}(t'-s)] &-\sin[\omega_{\bm q}(t'-s)] \\\sin[\omega_{\bm q}(t'-s)] &\cos[\omega_{\bm q}(t'-s)] 
    \end{bmatrix}
\end{equation*}
and    
\begin{equation*}
    \bar{\boldsymbol{M}}_{\bm q}(t'-s)=
    \begin{bmatrix}\sin[\omega_{\bm q}(t'-s)] &0 \\0 &-\sin[\omega_{\bm q}(t'-s)]
    \end{bmatrix},
\end{equation*}
we can express the mean-field action as
\begin{equation*}
    S_{\text{mf}}[\bm r(t')]=-\sum_{\bm q}\int_0^t \boldsymbol{X}^0\cdot\bar{\boldsymbol{L}}(t') \cdot\boldsymbol{g}(\boldsymbol{r'}) dt'.
\end{equation*}
and similarly the influence phase takes the form of
\begin{align*}
\Psi[\bm r(t),\bm r'(t)]=&
-\frac{1}{2}\int_0^t\int_0^t\sum_{\bm q}\\& \frac{1}{2}\bm v_{\bm q}(s)\cdot \bar{\boldsymbol{L}}_{\bm q}(t'-s)\cdot \bm v_{\bm q}(t')+ \\&
2i\bm v_{\bm q}(s)\cdot \bar{\boldsymbol{M}}_{\bm q}(t'-s)\Theta(s-t')\cdot \bm u_{\bm q}(t')dt'ds,
\end{align*}
where $\Theta$ is the Heaviside function.

Subsequently, by taking advantage of the Hubbard-Stratonovich transformation~\cite{stratonovich1958}, we can put forth a stochastic master equation for the reduced density matrix $\tilde{\rho}$ that yields the original density matrix $\rho_S$ when averaged over the established noise $W$, or formally $\rho_S = \langle \tilde{\rho}_S \rangle_W$. More specifically, we bring in extra noise variables $\bm \eta_{\bm q}$ and $\bm \nu_{\bm q}$
that are Gaussian-distributed with mean zero and covariance set as
\begin{subequations}
\label{eq:noise covariance}
\begin{align}
&\Big \langle \bm \eta^T_{\bm q}(s) \cdot \bm \eta_{\boldsymbol{r}}(t')\Big \rangle_W=\frac{1}{2} \bar{\boldsymbol{L}}_{\bm q}(t'-s)\delta_{\bm q,\boldsymbol{r}}\\
&\Big \langle \bm\eta^T_{\boldsymbol{q}}(s)\bm \cdot \bm \nu_{\bm r}(t')\Big \rangle_W=i\bar{\boldsymbol{M}}_{\bm q}(t'-s)\Theta(s-t')\delta_{\bm q,\boldsymbol{r}}\\
&\Big \langle \bm\nu^T_{\bm q}(s) \cdot \bm\nu_{\boldsymbol{r}}(t')\Big \rangle_W=0
\end{align}
\end{subequations}
In relation to this noise $W$, the influence phase takes the form of
\begin{align*}
\begin{split}
    &\Psi[\bm r(t),\bm r'(t)]=\\& - \ln \left[\left<\exp\left(i\int_0^t\sum_{\bm q} \bm{\eta}_{\bm q}(s)\cdot \bm{v}_{\bm q}(s)+\bm{\nu}_{\bm q}(s)\cdot\bm{u}_{\bm q}(s)\,ds \right)\right>_{W} \right].
\end{split}
\end{align*}
We can ergo found the master equation for the reduced density matrix $\tilde{\rho}_S$ associated with a single realization of the noise as
\begin{equation}
\label{eq:LvN}
\begin{split}
    i\frac{\partial}{\partial t}\tilde{\rho}_S&= \Big[\mathcal{H}_{S}+\mathcal{H}_{\text{mf}},\tilde\rho_S \Big]\\ &- \sum_{\boldsymbol{q}} \left( [\bm\eta_{\boldsymbol{q}}(t)\cdot \bm g_{\boldsymbol{q}},\tilde \rho_S]-\frac{1}{2}\{\bm\nu_{\boldsymbol{q}}(t)\cdot \bm g_{\boldsymbol{q}},\tilde \rho_S\} \right)
\end{split}
\end{equation}
where the mean-field Hamiltonian is defined by its action as
\begin{equation}
    \mathcal{H}_{\textrm{mf}} = \sum_{\bm q} \boldsymbol{X}^0\cdot\bar{\boldsymbol{L}} \cdot\boldsymbol{g}(\boldsymbol{r}).
\end{equation}

Notably, the derived master equation is valid beyond the weak system-bath coupling and the Markovian approximation. Nevertheless, we see that the mean-field approximation becomes more accurate in the weak-coupling limit, since the mean-field action is linear in the coupling whereas the influence phase is quadratic.

Furthermore, in contrast to Lindbladian alternatives~\cite{weiss2012quantum}, in our stochastic master equation the operators act only as $L\rho$ or $\rho L$, and not as $L\rho L$. Therefore, we can further decompose it into two independent Schrödinger equations with non-Hermitian Hamiltonians by applying the ansatz of $\tilde{\rho}_S=|\psi_{+}\rangle\langle\psi_{-}|$ where the states evolve as
\begin{equation}
\begin{split}
    i \frac{\partial}{\partial t}|\psi_{\pm}\rangle =\Bigg[\hat{H}_{S}+\hat{H}_{\textrm{mf}} - \sum_{\bm q} \Big(\bm \eta_{\bm q}(t)\cdot \bm g_{\bm q} \mp \frac{1}{2}\bm \nu_{\bm q}(t)\bm \cdot \bm g_{\bm q} \Big)\Bigg]|\psi_{\pm}\rangle.
\end{split}   
\label{eq:time evolution}
\end{equation}

Succinctly, to employ the quantum-acoustical master equation above, one first selects the system Hamiltonian $\mathcal{H}_S$ and set up the form of the interaction, i.e., the functions $\bm g_{\bm q}$. Then, the system state $|\psi_{\pm}\rangle$ and coherent state parameters $\bm X_{\bm q}$ are initialized. The evolution of the system  $|\psi_{\pm}(t)\rangle$ is determined according to Eq.~\ref{eq:time evolution} where the Gaussian noise is generated according to Eq.~\ref{eq:noise covariance}. The expectation value of any system observable $\hat O(t)$ can consequently be evaluated as 
\begin{equation}\label{eq:expectation_value_2}
    \langle\hat O \rangle (t) =\left<\langle\psi_-(t)|\hat O|\psi_+(t) \rangle\right>_W.
\end{equation}
even though we have traced away the bath and thus became agnostic to its time evolution, we are still able to recover the expected values of certain observables via the Ehrenfest theorem. For instance, the expected values of the position and momentum operators for a lattice mode $\boldsymbol{q}$ are given by
\begin{align}
\label{eq: Ehrenfest2}
    \frac{d}{dt} \boldsymbol{X}_{\boldsymbol{q}}(t) = \sigma_2\cdot\left(\omega_{\boldsymbol{q}}  \boldsymbol{X}_{\boldsymbol{q}}(t)  +\frac{1}{2}\langle \boldsymbol{g}_{\bm q} \rangle(t) \right)
\end{align}
where $\sigma_2$ is the Pauli matrix and the expectation values of $\bm g_{\bm q}(\bm r)$ are determined by Eq.~\ref{eq:expectation_value}. Therefore, the utilization of the Ehrenfest theorem leads to coupled linear differential equations mimicking the classical motion of a bath composed of harmonic oscillators.

\section{Quantum-acoustical model}

We further showcase the applicability of the formalistrange metal developed above within the context of the standard Fr{\"o}hlich model~\cite{Fro} where we have  
\begin{equation}
    \bm g_{\bm q}(\bm r)=
    \frac{E_d |\bm q|}{\sqrt{\rho\mathcal{V}\omega_{\bm q}}}
    \begin{bmatrix}
    \cos({\bm q\cdot\boldsymbol{r}} + \pi)\\
    \sin({\bm q\cdot\boldsymbol{r}}),
    \end{bmatrix}
\end{equation}
expressed with the wave vector $\mathbf{q}$ and frequency  $\omega_{\mathbf{q}}$ as well as the mass density $\rho$ and volume $\mathcal{V}$ of the lattice. The deformation potential constant $E_d$ characterizes the modulation of the electronic band energy due to lattice vibrations. Furthermore, the wave vector $\vert \mathbf{q} \vert \le q_D$ is restricted by the Debye wavenumber $q_D$, and we assume the linear dispersion $\omega_{\bm q}=v_s|\bm q|$ where $v_s$ is the speed of sound. We also focus on the electronic transport within the effective mass description. 

The mean-field component of the model then corresponds to the deformation potential $V_{\textrm{def}}$, which has previously been studied in Refs.~\cite{heller22, PhysRevLett.132, PNASpr, aydin_polaron_2025, zimmermann_rise_2024}. Since the deformation potential overall averages to zero, it is best characterized by its root-mean-square $\Delta V_\textrm{def}$. Assuming that the energy of charge carriers in a material can be attributed to the Fermi energy $E_F$, their dynamics with the underlying lattice can be roughly separated into two domains~\cite{zimmermann_rise_2024}:
\begin{align*}
\bar{K} = \frac{E_F}{\Delta V_\textrm{def}}
 \begin{cases}
	\gg 1  & \rightarrow \;  \textrm{Perturbative} \\
    \\
	\lesssim 1 \;  & \rightarrow \;  \textrm{Nonperturbative}
\end{cases}
\end{align*}
that equally classifies our electron-lattice coupling to be weak or strong, respectively.

For example, normal metals fall within the perturbative regime, while the class of compounds known as strange or bad metals exemplifies nonperturbative electron-lattice dynamics. In this work, we focus on Copper ($\bar{K} \gg 1$) and the prototypical strange metal Bi2212 ($\bar{K} \lesssim 1$). Additional computational details are provided in the Supplemental Material, based on experimental data and consistent with previous studies~\cite{heller22, PhysRevLett.132, PNASpr, aydin_polaron_2025, zimmermann_rise_2024}. However, we want to stress that the physics we find below transcends the material-specific constraints.

\bibliography{refs}

\providecommand{\noopsort}[1]{}\providecommand{\singleletter}[1]{#1}%
\begin{thebibliography}{10}

\bibitem{schr}
E.~Schr\"odinger.
\newblock Der stetige Übergang von der mikro- zur makromechanik.
\newblock {\em Naturwissenschaften}, 14:664--666, 1926.

\bibitem{heller22}
Donghwan Kim, Alhun Aydin, Alvar Daza, Kobra~N. Avanaki, Joonas Keski-Rahkonen, and Eric~J. Heller.
\newblock Coherent charge carrier dynamics in the presence of thermal lattice vibrations.
\newblock {\em Phys. Rev. B}, 106:054311, Aug 2022.

\bibitem{landau1948effective}
LD~Landau and SI~Pekar.
\newblock Effective mass of a polaron.
\newblock {\em Zh. Eksp. Teor. Fiz}, 18(5):419--423, 1948.

\bibitem{leopold2022landau}
Nikolai Leopold, David Mitrouskas, Simone Rademacher, Benjamin Schlein, and Robert Seiringer.
\newblock Landau--pekar equations and quantum fluctuations for the dynamics of a strongly coupled polaron.
\newblock {\em Pure and Applied Analysis}, 3(4):653--676, 2022.

\bibitem{PhysRevLett.132}
Joonas Keski-Rahkonen, Xiaoyu Ouyang, Shaobing Yuan, Anton~M. Graf, Alhun Aydin, and Eric~J. Heller.
\newblock Quantum-acoustical drude peak shift.
\newblock {\em Phys. Rev. Lett.}, 132:186303, May 2024.

\bibitem{Aydin24}
Alhun Aydin, Joonas Keski-Rahkonen, and Eric~J. Heller.
\newblock Quantum acoustics unravels planckian resistivity.
\newblock {\em Proceedings of the National Academy of Sciences}, 121(28):e2404853121, 2024.

\bibitem{aydin_polaron_2025}
Alhun Aydin, Joonas Keski-Rahkonen, Anton~M. Graf, Shaobing Yuan, Xiao-Yu Ouyang, Özgür~E. Müstecaplıoğlu, and Eric~J. Heller.
\newblock Polaron catastrophe within quantum acoustics.
\newblock {\em Proceedings of the National Academy of Sciences}, 122(23):e2426518122, June 2025.

\bibitem{kitaev2015}
Alexei Kitaev.
\newblock A simple model of quantum holography.
\newblock Talks at the KITP, April and May 2015.
\newblock \url{http://online.kitp.ucsb.edu/online/entangled15/kitaev/}.

\bibitem{rosenhaus2019}
Vladimir Rosenhaus.
\newblock An introduction to the syk model.
\newblock {\em Journal of Physics A: Mathematical and Theoretical}, 52:323001, 2019.

\bibitem{ashcroft1976solid}
N.W. Ashcroft and N.D. Mermin.
\newblock {\em Solid State Physics}.
\newblock HRW international editions. Holt, Rinehart and Winston, 1976.

\bibitem{glauber1963}
Roy~J. Glauber.
\newblock Coherent and {Incoherent} {States} of the {Radiation} {Field}.
\newblock {\em Physical Review}, 131(6):2766--2788, September 1963.

\bibitem{ShockleyBardeen}
W.~Shockley and J.~Bardeen.
\newblock Energy bands and mobilities in monatomic semiconductors.
\newblock {\em Phys. Rev.}, 77:407--408, Feb 1950.

\bibitem{sun_lorenz_2024}
Fei Sun, Simli Mishra, Ulrike Stockert, Ramzy Daou, Naoki Kikugawa, Robin~S. Perry, Elena Hassinger, Sean~A. Hartnoll, Andrew~P. Mackenzie, and Veronika Sunko.
\newblock The {Lorenz} ratio as a guide to scattering contributions to transport in strongly correlated metals.
\newblock {\em Proceedings of the National Academy of Sciences}, 121(35):e2318159121, August 2024.
\newblock Publisher: Proceedings of the National Academy of Sciences.

\bibitem{linehan_listening_2024}
Ryan Linehan, Tanner Trickle, Christopher~R. Conner, Sohitri Ghosh, Tongyan Lin, Mukul Sholapurkar, and Andrew~N. Cleland.
\newblock Listening {For} {New} {Physics} {With} {Quantum} {Acoustics}, October 2024.
\newblock arXiv:2410.17308.

\bibitem{9880558}
Michael Choquer, Matthias Weiß, Emeline D.~S. Nysten, Michelle Lienhart, PaweŁ Machnikowski, Daniel Wigger, Hubert~J. Krenner, and Galan Moody.
\newblock Quantum control of optically active artificial atoms with surface acoustic waves.
\newblock {\em IEEE Transactions on Quantum Engineering}, 3:1--17, 2022.

\bibitem{hellerkim}
Eric~J. Heller and Donghwan Kim.
\newblock Schr\"{o}dinger correspondence applied to crystals.
\newblock {\em The Journal of Physical Chemistry A}, 123(20):4379--4388, 2019.
\newblock PMID: 30892041.

\bibitem{Many-Particle}
G.~D. Mahan.
\newblock {\em Many-Particle Physics}.
\newblock Plenum, New York, third edition, 2000.

\bibitem{frohlich1954electrons}
Herbert Fr{\"o}hlich.
\newblock Electrons in lattice fields.
\newblock {\em Advances in Physics}, 3(11):325--361, 1954.

\bibitem{scully1997quantum}
M.O. Scully and M.S. Zubairy.
\newblock {\em Quantum Optics}.
\newblock Cambridge University Press, 1997.

\bibitem{walls2007quantum}
D.F. Walls and G.J. Milburn.
\newblock {\em Quantum Optics}.
\newblock Springer Berlin Heidelberg, 2007.

\bibitem{leopold2021derivation}
Nikolai Leopold, David Mitrouskas, and Robert Seiringer.
\newblock Derivation of the landau--pekar equations in a many-body mean-field limit.
\newblock {\em Archive for Rational Mechanics and Analysis}, 240(1):383--417, 2021.

\bibitem{pekar1946local}
Solomon Pekar.
\newblock Local quantum states of electrons in an ideal ion crystal.
\newblock {\em Zh. Eksp. Teor. Fiz}, 16(4):341--348, 1946.

\bibitem{frank_zhou_2017}
Rupert~L. Frank and Gang Zhou.
\newblock Derivation of an effective evolution equation for a strongly coupled polaron.
\newblock {\em Analysis \& PDE}, 10(2):379–422, 2017.
\newblock Funding by NSF.

\bibitem{HBT1}
R.~Hanbury Brown and R.~Q. Twiss.
\newblock Interferometry of the intensity fluctuations in light. i. basic theory: The correlation between photons in coherent beams of radiation.
\newblock {\em Proceedings of the Royal Society of London. Series A, Mathematical and Physical Sciences}, 242(1230):300--324, 1957.

\bibitem{PNASpr}
Alhun Aydin, Joonas Keski-Rahkonen, and Eric~J. Heller.
\newblock Quantum acoustics unravels {Planckian} resistivity.
\newblock {\em Proceedings of the National Academy of Sciences}, 121(28):e2404853121, July 2024.
\newblock Publisher: Proceedings of the National Academy of Sciences.

\bibitem{FeynmanInfluence}
R.P Feynman and F.L Vernon.
\newblock The theory of a general quantum system interacting with a linear dissipative system.
\newblock {\em Annals of physics}, 24:118--173, 1963.

\bibitem{STOCKBURGER2001249}
Jürgen~T. Stockburger and Hermann Grabert.
\newblock Non-markovian quantum state diffusion.
\newblock {\em Chemical Physics}, 268(1):249--256, 2001.

\bibitem{joost}
Joost de~Nijs, Anton~M. Graf, Joonas Keski-Rahkonen, and Eric~J. Heller.
\newblock Path-integral road to quantum acoustics.
\newblock {\em arXiv preprint arXiv:xxxxxx}, 2025.

\bibitem{weiss2012quantum}
Ulrich Weiss.
\newblock {\em Quantum dissipative systems}.
\newblock World Scientific, 2012.

\bibitem{PLLTransformation}
T.~D. Lee, F.~E. Low, and D.~Pines.
\newblock The motion of slow electrons in a polar crystal.
\newblock {\em Phys. Rev.}, 90:297--302, Apr 1953.

\bibitem{kubo2012statistical}
Ryogo Kubo, Morikazu Toda, and Natsuki Hashitsume.
\newblock {\em Statistical physics II: nonequilibrium statistical mechanics}, volume~31.
\newblock Springer Science \& Business Media, 2012.

\bibitem{grissonnanche_linear-temperature_2021}
Gaël Grissonnanche, Yawen Fang, Anaëlle Legros, Simon Verret, Francis Laliberté, Clément Collignon, Jianshi Zhou, David Graf, Paul~A. Goddard, Louis Taillefer, and B.~J. Ramshaw.
\newblock Linear-in temperature resistivity from an isotropic {Planckian} scattering rate.
\newblock {\em Nature}, 595(7869):667--672, July 2021.

\bibitem{Hartnoll2015}
Sean~A. Hartnoll.
\newblock Theory of universal incoherent metallic transport.
\newblock {\em Nature Physics}, 11(1):54--61, Jan 2015.

\bibitem{Nussinov_ann.phys_443_168970_2022}
Zohar Nussinov and Saurish Chakrabarty.
\newblock Exact universal chaos, speed limit, acceleration, planckian transport coefficient,“collapse” to equilibrium, and other bounds in thermal quantum systems.
\newblock {\em Ann. Phys.}, 443:168970, 2022.

\bibitem{NUSSINOV2022168970}
Zohar Nussinov and Saurish Chakrabarty.
\newblock Exact universal chaos, speed limit, acceleration, planckian transport coefficient, “collapse” to equilibrium, and other bounds in thermal quantum systems.
\newblock {\em Annals of Physics}, 443:168970, 2022.

\bibitem{ghost}
Yubo Zhang, Anton~M. Graf, Alhun Aydin, Joonas Keski-Rahkonen, and Eric~J. Heller.
\newblock Planckian diffusion: the ghost of anderson localization.
\newblock {\em arXiv preprint arXiv:2411.18768}, 2024.

\bibitem{Zaanen_Sci.post_6_061_2019}
Jan Zaanen.
\newblock Planckian dissipation, minimal viscosity and the transport in cuprate strange metals.
\newblock {\em SciPost Physics}, 6(5):061, 2019.

\bibitem{Hartnoll_rev.mod.phys_94_041002_2022}
Sean~A Hartnoll and Andrew~P Mackenzie.
\newblock Colloquium: Planckian dissipation in metals.
\newblock {\em Rev. Mod. Phys.}, 94(4):041002, 2022.

\bibitem{Mousatov_nat.phys_16_579_2020}
Connie~H Mousatov and Sean~A Hartnoll.
\newblock On the planckian bound for heat diffusion in insulators.
\newblock {\em Nat. Phys.}, 16(5):579--584, 2020.

\bibitem{Zhang_pnas_116_19869_2019}
Jiecheng Zhang, Erik~D Kountz, Kamran Behnia, and Aharon Kapitulnik.
\newblock Thermalization and possible signatures of quantum chaos in complex crystalline materials.
\newblock {\em Proc. Natl. Acad. Sci.}, 116(40):19869--19874, 2019.

\bibitem{Lucas_phys.rev.lett_122_216601_2019}
Andrew Lucas.
\newblock Operator size at finite temperature and planckian bounds on quantum dynamics.
\newblock {\em Phys. Rev. Lett.}, 122(21):216601, 2019.

\bibitem{Gleis_phys.rev.lett_134_106501_2025}
Andreas Gleis, Seung-Sup~B Lee, Gabriel Kotliar, and Jan Von~Delft.
\newblock Dynamical scaling and planckian dissipation due to heavy-fermion quantum criticality.
\newblock {\em Phys. Rev. Lett.}, 134(10):106501, 2025.

\bibitem{adams_localization_1987}
Philip~W. Adams and Mikko~A. Paalanen.
\newblock Localization in a {Nondegenerate} {Two}-{Dimensional} {Electron} {Gas}.
\newblock {\em Physical Review Letters}, 58(20):2106--2109, May 1987.

\bibitem{adams_conductivity_1992}
P.~W. Adams.
\newblock The conductivity and mobility of {2D} nondegenerate electrons in the strong localization regime.
\newblock {\em Surface Science}, 263(1):663--667, February 1992.

\bibitem{Edwards_1972}
J~T Edwards and D~J Thouless.
\newblock Numerical studies of localization in disordered systems.
\newblock {\em Journal of Physics C: Solid State Physics}, 5(8):807--820, apr 1972.

\bibitem{infrared}
S.~Fratini and S.~Ciuchi.
\newblock {Displaced Drude peak and bad metal from the interaction with slow fluctuations.}
\newblock {\em SciPost Phys.}, 11:39, 2021.

\bibitem{prlDDP}
J.~Keski-Rahkonen, X.~Y. Ouyang, S.~Yuan, A.~M. Graf, A.~Aydin, and E.~J. Heller.
\newblock Quantum-acoustical drude peak shift.
\newblock {\em Physical Review Letters}, 2023.

\bibitem{strangemetal1}
J.~A.~N. Bruin, H.~Sakai, R.~S. Perry, and A.~P. Mackenzie.
\newblock Similarity of scattering rates in metals showing \textit{T}-linear resistivity.
\newblock {\em Science}, 339(6121):804--807, 2013.

\bibitem{planckianmetals}
Sean~A. Hartnoll and Andrew~P. Mackenzie.
\newblock Colloquium: Planckian dissipation in metals.
\newblock {\em Rev. Mod. Phys.}, 94:041002, Nov 2022.

\bibitem{Grissonnanche2021}
Ga{\"e}l Grissonnanche, Yawen Fang, Ana{\"e}lle Legros, Simon Verret, Francis Lalibert{\'e}, Cl{\'e}ment Collignon, Jianshi Zhou, David Graf, Paul~A. Goddard, Louis Taillefer, and B.~J. Ramshaw.
\newblock Linear-in temperature resistivity from an isotropic planckian scattering rate.
\newblock {\em Nature}, 595(7869):667--672, Jul 2021.

\bibitem{stratonovich1958}
R.~L. Stratonovich.
\newblock On a method of calculating quantum distribution functions.
\newblock {\em Soviet Physics Doklady}, 2:416, 1958.

\bibitem{Fro}
H.~Fr\"olich.
\newblock {\em Theorie der Metalle}.
\newblock Springer, 1939.

\bibitem{zimmermann_rise_2024}
Yoel Zimmermann, Joonas Keski-Rahkonen, Anton~M Graf, and Eric~J Heller.
\newblock Rise and fall of anderson localization by lattice vibrations: {A} time-dependent machine learning approach.
\newblock {\em Entropy}, 26(7):552, 2024.
\newblock Publisher: MDPI.

\end{thebibliography}
\bibliographystyle{unsrt}

\end{document}